\documentclass[%
 reprint,
superscriptaddress,
nofootinbib,
 amsmath,amssymb,
 aps,
]{revtex4-1}
\usepackage{array}
\usepackage{graphicx}
\usepackage{dcolumn}
\usepackage{bm}
\usepackage{subcaption}
\usepackage{color}
\usepackage{hyperref}

\newcommand{\beq}{\begin{equation}}
\newcommand{\eeq}{\end{equation}}

\makeatletter
\newcommand{\thickhline}{%
    \noalign {\ifnum 0=`}\fi \hrule height 1pt
    \futurelet \reserved@a \@xhline
}
\newcolumntype{M}[1]{>{\centering\arraybackslash}m{#1}}
\newcolumntype{'}{@{\hskip\tabcolsep\vrule width 1pt\hskip\tabcolsep}}
\makeatother

\begin{document}


\title{DeepXS: Fast approximation of MSSM electroweak cross sections at NLO}

\author{Sydney Otten}\email{Sydney.Otten@ru.nl}
\affiliation{Institute for Mathematics, Astro- and Particle Physics IMAPP\\Radboud Universiteit, Nijmegen, The Netherlands}
\affiliation{GRAPPA, University of Amsterdam, The Netherlands}
\author{Krzysztof Rolbiecki}%
 \email{Krzysztof.Rolbiecki@fuw.edu.pl}
\affiliation{
 Faculty of Physics, University of Warsaw, Poland
}
\author{Sascha Caron}
\affiliation{Institute for Mathematics, Astro- and Particle Physics IMAPP\\Radboud Universiteit, Nijmegen, The Netherlands}
\affiliation{Nikhef, Amsterdam, The Netherlands}
\author{Jong-Soo Kim}
\affiliation{
 Mandelstam Institute for Theoretical Physics
}
\affiliation{
 National Institute for Theoretical Physics\\
 University of the Witwatersrand, Johannesburg, South Africa
}
\author{Roberto Ruiz de Austri}
\affiliation{
 Instituto de Fisica Corpuscular, IFIC-UV/CSIC\\
 University of Valencia, Spain
}
\author{Jamie Tattersall}
\affiliation{
 Institute for Theoretical Particle Physics and Cosmology\\
 RWTH Aachen University, Germany
}
\affiliation{ESR Labs, Munich, Germany}
\date{\today}

\begin{abstract}
We present a deep 
learning solution to the prediction of particle production cross sections over a complicated, high-dimensional parameter space. We demonstrate the applicability by providing state-of-the-art predictions for the production of charginos and neutralinos at the Large Hadron Collider (LHC) at the next-to-leading order in the phenomenological MSSM-19 and explicitly demonstrate the performance for $pp\to\tilde{\chi}^+_1\tilde{\chi}^-_1,$ $\tilde{\chi}^0_2\tilde{\chi}^0_2$ and $\tilde{\chi}^0_2\tilde{\chi}^\pm_1$ as a proof of concept which will be extended to all SUSY electroweak pairs. We obtain errors that are lower than the uncertainty from scale and parton distribution functions with mean absolute percentage errors of well below $0.5\,\%$ allowing a safe inference at the next-to-leading order with inference times that improve the Monte Carlo integration procedures that have been available so far by a factor of $\mathcal{O}(10^7)$ from $\mathcal{O}(\rm{min})$ to $\mathcal{O}(\mu\rm{s})$ per evaluation.
\end{abstract}

\pacs{Valid PACS appear here}
 \maketitle

\section{\label{sec:intro}Introduction}

Dimensionality persists to be a curse for everyone that seeks the needle in a complex haystack. Despite all the achievements from data science so far, physicists often resort to simplified, lower-dimensional models to obtain a tractable problem~\cite{Abdallah:2015ter, Athron:2017yua}. This strategy prevents the scientific community from utilising all the available information to pin down the laws of nature. To overcome this very general issue, we investigate deep learning techniques as a potential solution motivated by the successful application of neural networks to cross sections with a four-dimensional parameter space~\cite{Bertone:2017adx}. We find positive results for cross sections that depend on a 19-dimensional parameter space with highly complex structures.

One of the most widely studied beyond the Standard Model (BSM) theories remains supersymmetry (SUSY) \cite{Drees:2004jm,Haber:1984rc,Martin:1997ns}. The ever increasing sophistication of experimental analyses requires that theoretical tools match the precision requirements set by experiments. One of the key requirements is performing cross section calculations of BSM processes at least at the next-to-leading order (NLO) accuracy. This goal has been gradually reached over many years, for particles produced both by strong and weak interactions, and the current state-of-the-art calculations also include resummed higher order corrections~\cite{Fuks:2012qx,Beenakker:2014sma}. Currently, for most applications it is possible and sufficient to calculate the production cross section at the next-to-leading-log approximation. However, such calculations are typically time consuming, e.g.\ it takes about three minutes for the computer program \texttt{Prospino}~\cite{Beenakker:1999xh} to calculate the chargino pair production cross section, $p p \to \tilde{\chi}^+_1 \tilde{\chi}^-_1$, at NLO. The computational time for \texttt{Resummino}~\cite{Fuks:2013vua} is similar at NLO but taking into account higher order corrections increases the time consumption 20 fold.

Many applications, for example global scans of the multi-dimensional parameter space of the Minimal Supersymmetric Standard Model (MSSM), see e.g.~\cite{Bechtle:2012zk, Strege:2014ija, Bagnaschi:2017tru, Athron:2018vxy,Bertone:2015tza,Kim:2016rsd}, demand a much faster method for the computation of NLO cross sections. In the case of strongly produced SUSY particles, this problem is addressed by the computer program \texttt{NNLLfast}~\cite{Beenakker:2016lwe} that offers an approximation of relevant cross sections within a fraction of a second at the next-to-next-to-leading logarithmic accuracy. 

\begin{figure}
\includegraphics[width=0.49\textwidth]{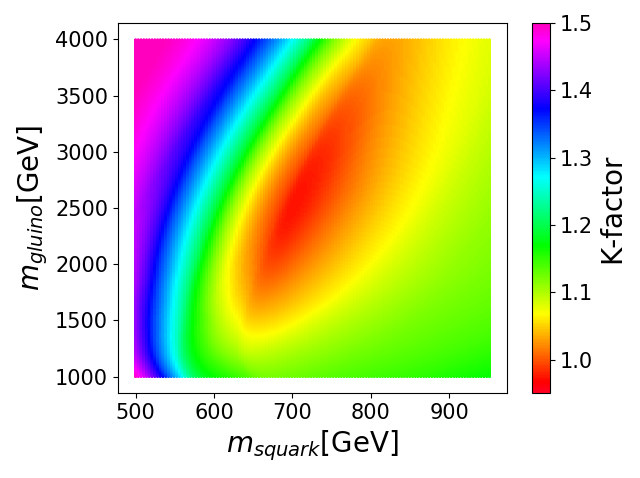}
\caption{A temperature plot for the $K$-factor in the wino scenario, predicted by a neural network, for $\tilde{\chi}^0_2\tilde{\chi}^+_1$ in the $m_{\tilde{q}}/m_{\tilde{g}}$ plane, already showing a non-trivial $K$-factor landscape for two free parameters. The electroweakino masses are set to 400\,GeV. }
\label{fig:sqgl}
\end{figure}

In this paper we present a novel approach that enables a fast approximation of cross sections in a high-dimensional parameter space and as an example demonstrate the applicability for chargino and neutralino~\footnote{Charginos and neutralinos are supersymmetric partners of the Standard Model gauge and Higgs bosons. In the following we will use the umbrella term ``electroweakinos''.} production cross sections at NLO accuracy in the phenomenological MSSM-19 (pMSSM-19), where 19 denotes the number of model parameters. We employ a machine learning technique to approximate the result from the cross sections calculated using \texttt{Prospino}. While the task might appear straightforward, there are several challenges that one has to solve to obtain a tool that provides both speed and high accuracy. Firstly, the cross sections span over up to 13 orders of magnitude, depending on the electroweakino masses and couplings. Secondly, the electroweakino sector is parametrized by four independent parameters in the SUSY Lagrangian and, in addition, the cross sections depend on the other SUSY particles, either at the tree-level (squarks) or at the loop level (gluinos). Thus, we need to construct representations for complicated functions whose effective parameter space that one has to cover can have up to 14 dimensions. A temperature plot showing the non-trivial $K$-factor landscape in only two of these dimensions is shown in Fig.~\ref{fig:sqgl} for $\tilde{\chi}^0_2\tilde{\chi}^+_1$. Here, we focus on the four most relevant processes at the LHC, i.e.\ the production of chargino pairs, $\tilde{\chi}^+_1 \tilde{\chi}^-_1$, neutralino pairs, $\tilde{\chi}^0_2 \tilde{\chi}^0_2$, and associated production of a chargino and a neutralino, $\tilde{\chi}^0_2 \tilde{\chi}^\pm_1$. The approach that we present here can be extended to other electroweak processes and models, e.g.\ the next-to-MSSM scenarios~\cite{Ellwanger:2009dp}.  

\section{\label{sec:DL}Methodology}

In order to develop a code that can predict values of an otherwise computationally expensive function fast and reliably---the cross section in our case---we take the following approach. First we calculate values of the function at a large number of points, $10^7$ samples at the leading order (LO) and $\mathcal{O}(10^5)$ samples at NLO. The points are sampled randomly in a high-dimensional parameter space in given ranges. The data is then used to train a customised artificial neural network (ANN), which is adopting deep learning techniques, stacking and an iterative ANN-based point selection procedure that picks points from a labeled pool of samples. The properly trained model is then able to provide accurate predictions of the cross section at a given parameter point. The performance of the resulting ANN is tested with $10^4$ samples that the deep network has never seen during training.     

\subsection{Data Generation}

The pMSSM-19 parameters are sampled with a flat prior within the ranges given in Table~2 of Ref.~\cite{Aad:2015baa}. 
Since sleptons do not affect the actual calculation of the cross section at any stage they are assumed to be mass degenerate between left and right-handed states for the first and second generations. These parameter sets are then passed to \verb+SPheno 3.38+~\cite{Porod:2003um,Porod:2011nf} to calculate the spectrum with default settings which is then fed into \verb+Prospino 2.1+, which calculates the cross section using CTEQ6 parton distribution functions (PDFs)~\cite{Pumplin:2002vw,Nadolsky:2008zw}. 

The NLO cross section can be written as a product of the $K$-factor and the LO cross section: 
\beq 
\sigma_{\rm{NLO}}=K\cdot\sigma_{\rm{LO}}.
\eeq 
Since most difficulties in the structure already appear at the leading order and the $K$-factor is a slowly varying function of the input parameters, we construct the NLO prediction by multiplying the predictions of the LO and $K$-factor regressors. This significantly decreases the computational cost by reducing the amount of necessary NLO data by two orders of magnitude. 

The relevant masses and mixing angles from the spectrum with the corresponding LO cross section and/or $K$-factor are systematically collected so that they can be used to optimise an ANN implementation as training and validation data. We preselect the samples such that $\tilde{\chi}_1^0$ is the lightest supersymmetric particle (LSP), there are no tachyons in the spectrum and no potential numerical problems occurred while running \verb+SPheno+. For all LO cross sections, we have created $10^7$ samples. For the $K$-factors, the number of generated samples varies between $1$--$6\times 10^5$. 

Because the starting point of the data generation is the pMSSM-19 parameter space, we must cover it appropriately and therefore we are confronted with the curse of dimensionality. To tackle this, we manually restrict the parameter space by excluding all cross sections that are not relevant. By exploiting the fact that the number of events $N$ at the LHC is equal to the product of the integrated luminosity and the cross section: 
\beq 
N=L_{\rm{int}}\cdot\sigma,\label{eq:NLsigma} 
\eeq 
and assuming the final integrated luminosity of the LHC to be $L_{\rm{int}}=3000\,\rm{fb}^{-1}$, we can derive a lower bound for the cross section by demanding at least one event in the life-time of the LHC. The resulting lower bound is $\sigma_{\rm{min}}= 3.3\cdot 10^{-7}\,\rm{pb}$. 

The generated data is then processed by a deep learning pipeline that utilises the ANN-based point selection (NNPS) and stacking, i.e.\ a manually implemented logical connection of different, eventually specialised predictors for the same task. Since we assume that most use-cases will run a spectrum calculator anyway and \verb+SPheno+ barely consumes computational capacity, we only create a neural network representation for the mapping from the masses and mixing angles to the LO cross section and $K$-factor.

\subsection{Optimising the Representations}

The deep learning techniques used here are employed via artificial neural networks implemented with \verb+Keras+~\cite{chollet2015keras} and a \verb+Tensorflow+~\cite{Abadi:2016:TSL:3026877.3026899} backend that were trained on a GPU using \verb+CUDA+~\cite{Nickolls:2008:SPP:1365490.1365500} and \verb+cuDNN+~\cite{cuDNN}. The pre- and post-processing of the input data together with the neural network architecture and the machine learning model parameters form the technical realization of the deep learning representation of the function $\sigma=\sigma(\textrm{pMSSM-19})$.  

The input of the neural networks is taken from the \verb+SPheno+ output and consists of the electroweakino and squark masses for the LO cross sections, and gluino mass for the $K$-factor, as well as the relevant chargino and neutralino mixing matrix entries, and is preprocessed via the $z$-score normalisation: the inputs $x_i$ are transformed into $x_i'=\frac{x_i-\mu(x)}{\sigma_{\rm{sd}}(x)}$, where $\mu (x)$ and $\sigma_{\rm{sd}}(x)$ are the mean and standard deviation of $x$. Whenever deemed useful, expert knowledge was applied and high-level features were formed, e.g.\ for the $K$-factor prediction, the mean of the squark masses was used, which corresponds to the calculation method employed in \verb+Prospino+. 

An ANN is a collection of artificial neurons, along whose connections an input is propagated. During the propagation the input is transformed depending on the network architecture and the machine learning model parameter set $\theta$ characterising the ANN. The output is an estimate of the function value for the given input parameters. The set $\theta$ is initially drawn from a random distribution and learned via updates from a stochastic gradient descent-like optimisation algorithm that minimises a loss function which measures the deviation between model predictions and true (known) cross sections. In our case, the loss function is the mean absolute percentage error
\beq
\mathrm{MAPE} = \frac{1}{N}\sum_{i=0}^N\left|\frac{y_{\rm{true},i}-y_{\rm{pred},i}}{y_{\rm{true},i}}\right|,
\eeq
which is minimised. The chosen optimiser is \verb+ADAM+~\cite{adam} with default parameters except a learning rate scheduling with initial and final learning rates $\alpha_i$ and $\alpha_f$ combined with \verb+EarlyStopping+~\cite{earlystopping}. When an iteration has ended, i.e.\ when either the pre-defined maximum number of epochs is reached or \verb+EarlyStopping+ terminates the process, the learning rate is divided by 2, the weights giving the best validation loss so far are loaded into the architecture and the optimisation continues until $\alpha_f$ is reached.
 
Due to the high computational cost of hyperparameter scans, i.e. many hours to days for a single hyperparameter point and the fact that the hyperparameter space is high-dimensional with a mixture of integer and continuous dimensions, we choose a heuristic approach to determine the hyperparameters.
For different processes of electroweakino production we therefore adapt different techniques to achieve MAPEs below 0.5\,\% and maximum errors of below 10\,\%.

The $\sigma_{\rm{LO}}$ input is propagated through eight hidden layers with 100 neurons each and the \verb+selu+~\cite{selu} activation function, while for the $K$-factors only 32 neurons per layer are used. The inputs are labeled with the corresponding cross section, in most cases pre-processed with a shifted logarithm such that for the input $x_i$ its label is given by 
\beq
y_i'=-\min(\log(\mathbf{\sigma}))+\log(\sigma_i)
\eeq
or for the $K$-factor divided by 2 or 4, depending on the pair. The loss function is the MAPE for the $K$-factors and a modification thereof for the LO cross sections that takes into account the pre-processing. In the default setup, the MAPE would minimise the error on $\log (\sigma)$ and it would result in sub-optimal performance. Therefore, several custom loss functions have been implemented that are constructed such that the loss function explicitly minimises the MAPE of the original values. The final data for the LO cross sections has been trained for 150 epochs and 7--10 iterations with $\alpha_i=0.0008$, a patience of 50 and a batch-size of 120. For the $K$-factors we used larger batch-sizes of between 512 and 1024 and a larger number of epochs of up to 250 per iteration. For $\tilde{\chi}^+_1\tilde{\chi}^-_1$ and $\tilde{\chi}^0_2\tilde{\chi}^0_2$ we  take all of the randomly generated samples and train deep networks for the LO and the $K$-factors.

\begin{figure*}
\begin{subfigure}{0.245\textwidth}
\includegraphics[width=\textwidth]{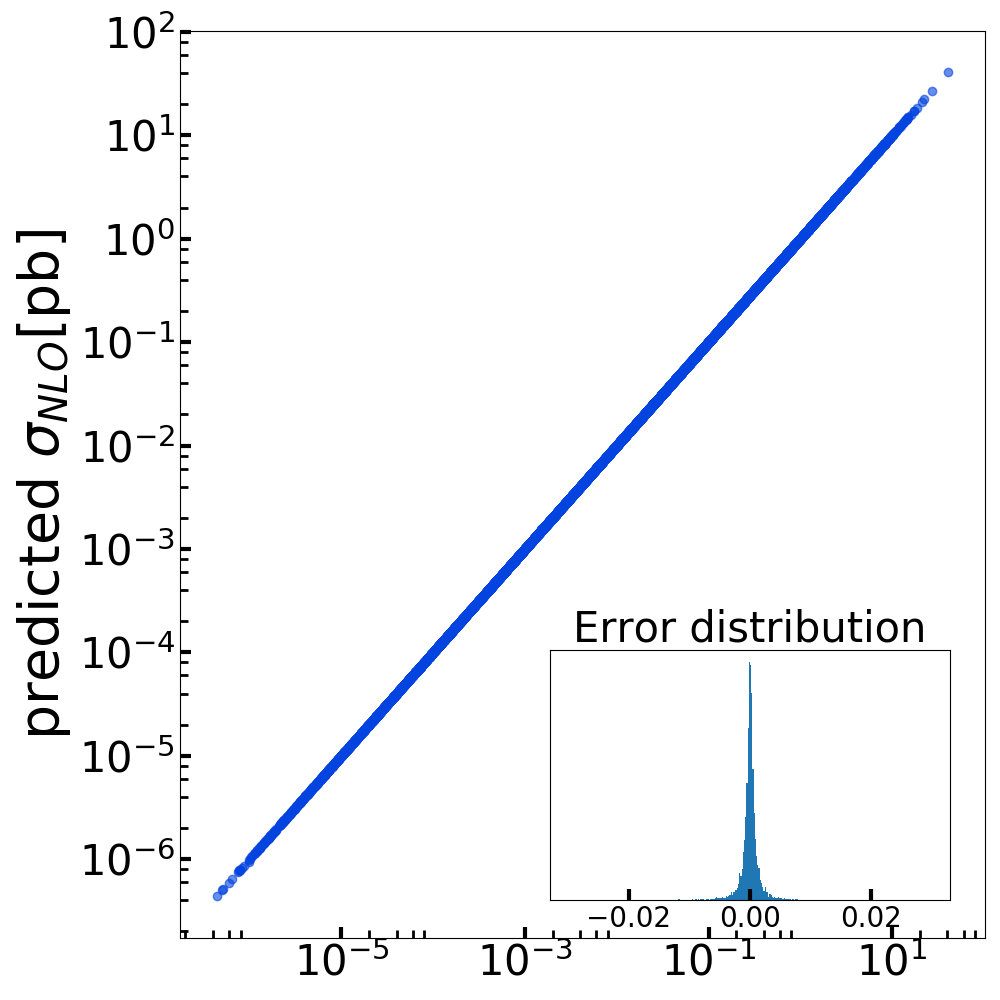}
\includegraphics[width=\textwidth]{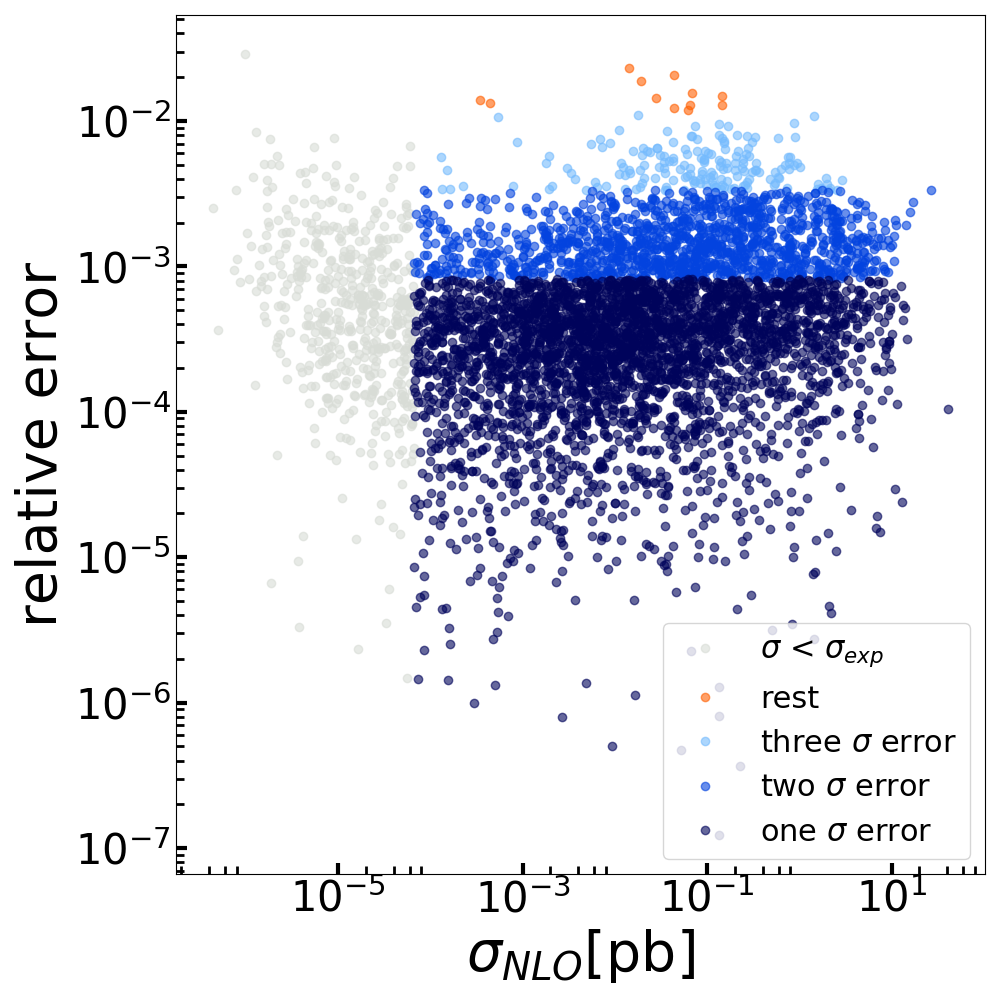}
\subcaption{$\tilde{\chi}^+_1\tilde{\chi}^-_1$}
\end{subfigure}
\begin{subfigure}{0.245\textwidth}
\includegraphics[width=\textwidth]{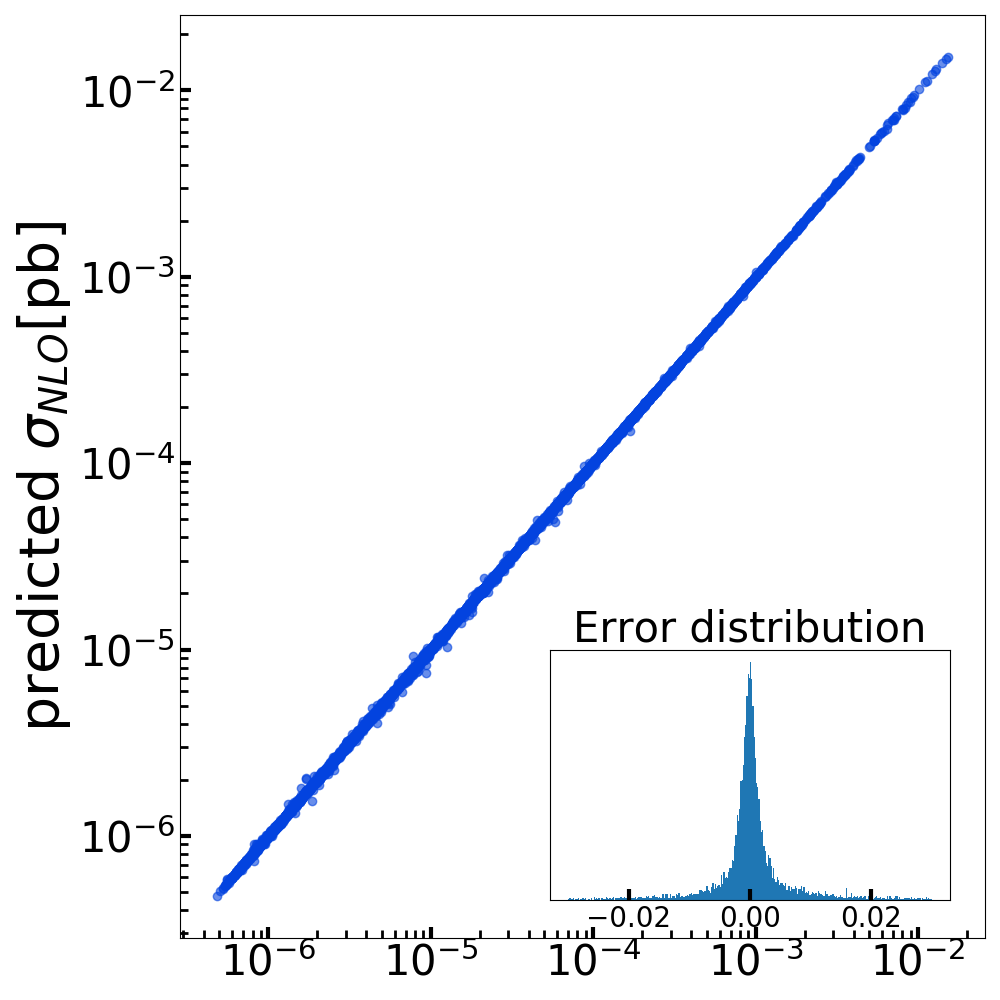}
\includegraphics[width=\textwidth]{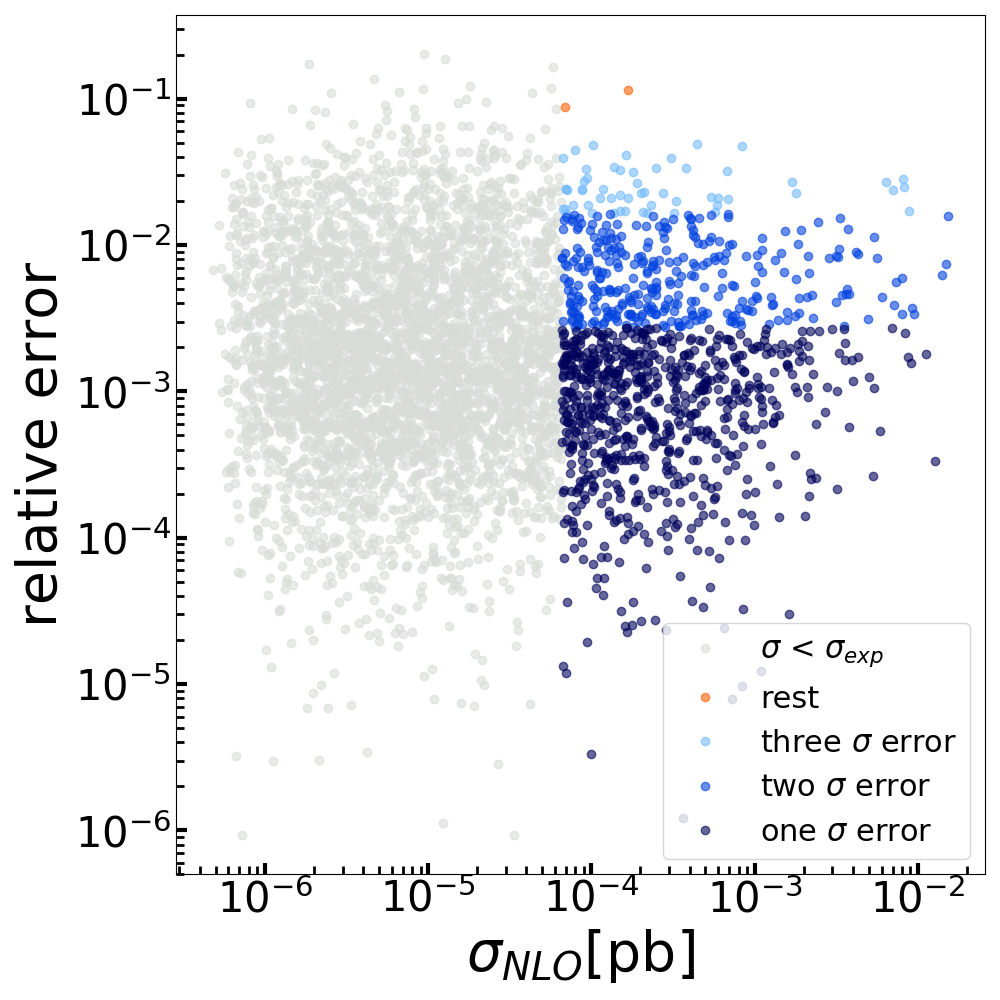}
\subcaption{$\tilde{\chi}^0_2\tilde{\chi}^0_2$}
\end{subfigure}
\begin{subfigure}{0.245\textwidth}
\includegraphics[width=\textwidth]{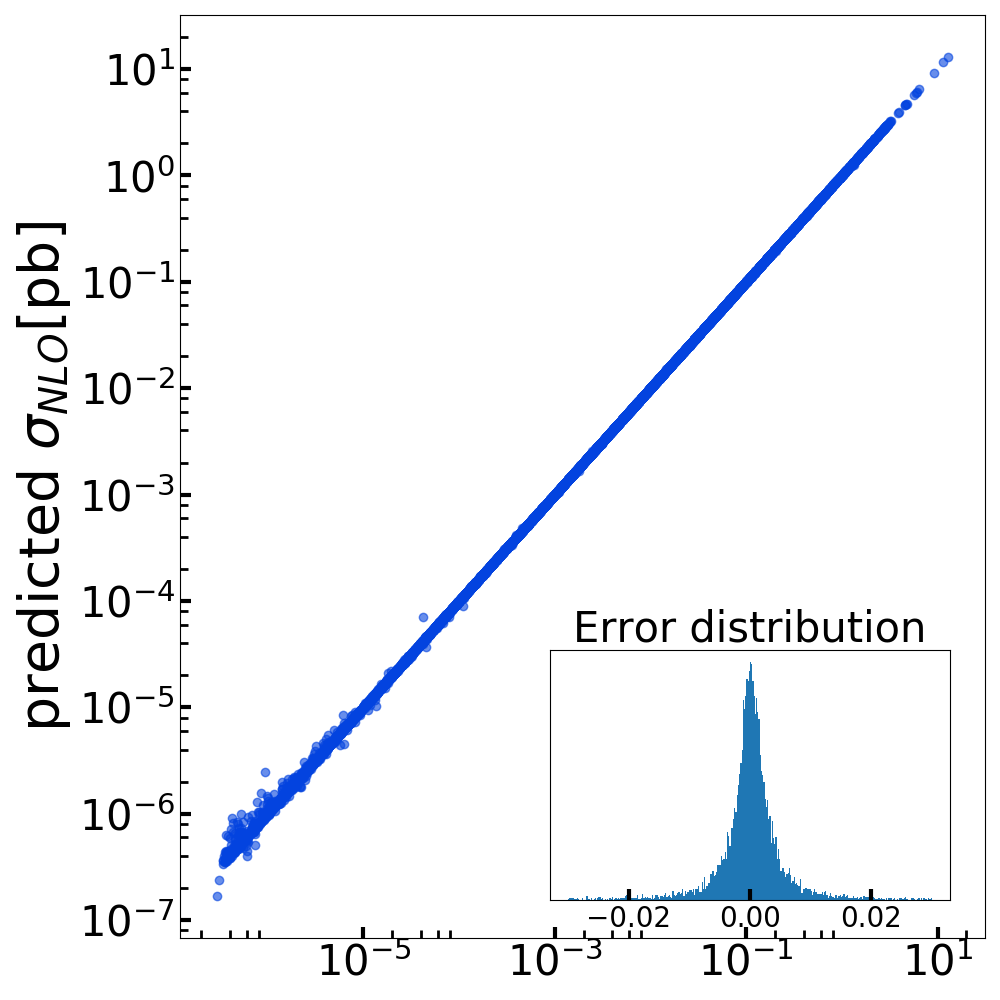}
\includegraphics[width=\textwidth]{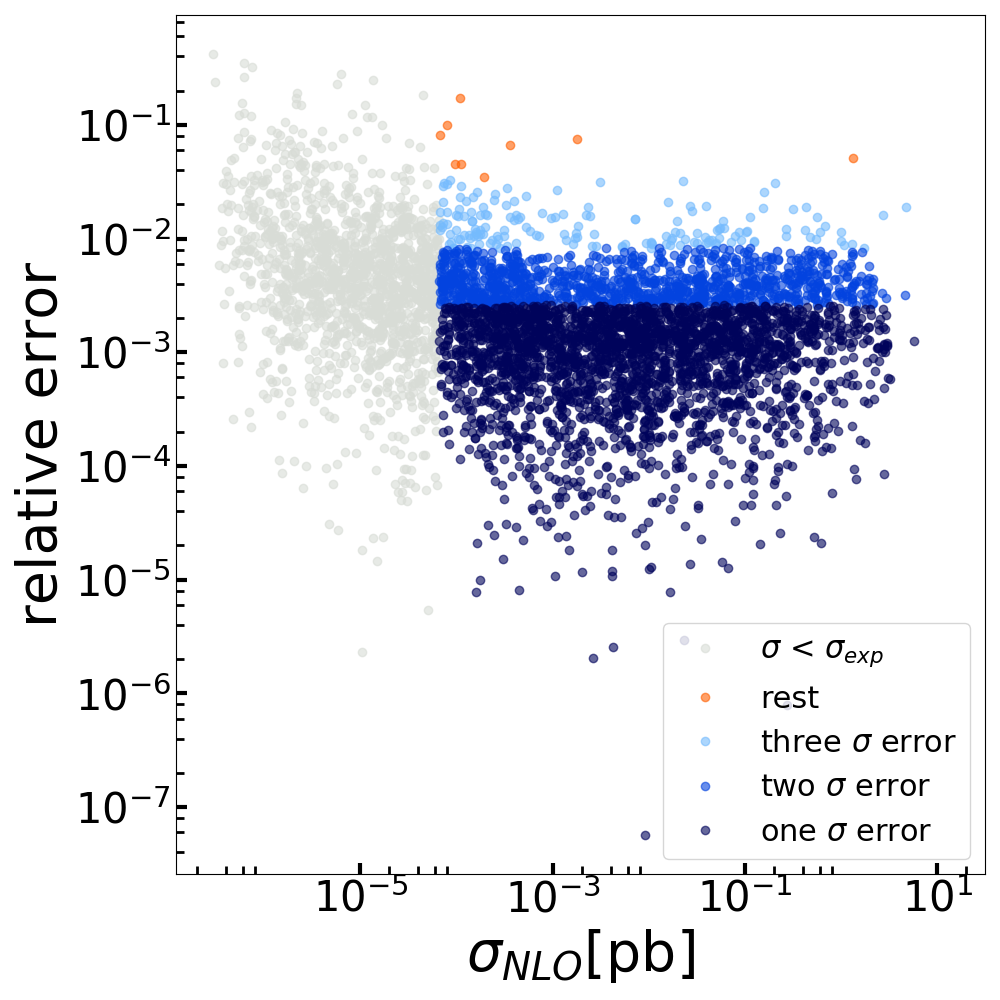}
\subcaption{$\tilde{\chi}^0_2\tilde{\chi}^+_1$}
\end{subfigure}
\begin{subfigure}{0.245\textwidth}
\includegraphics[width=\textwidth]{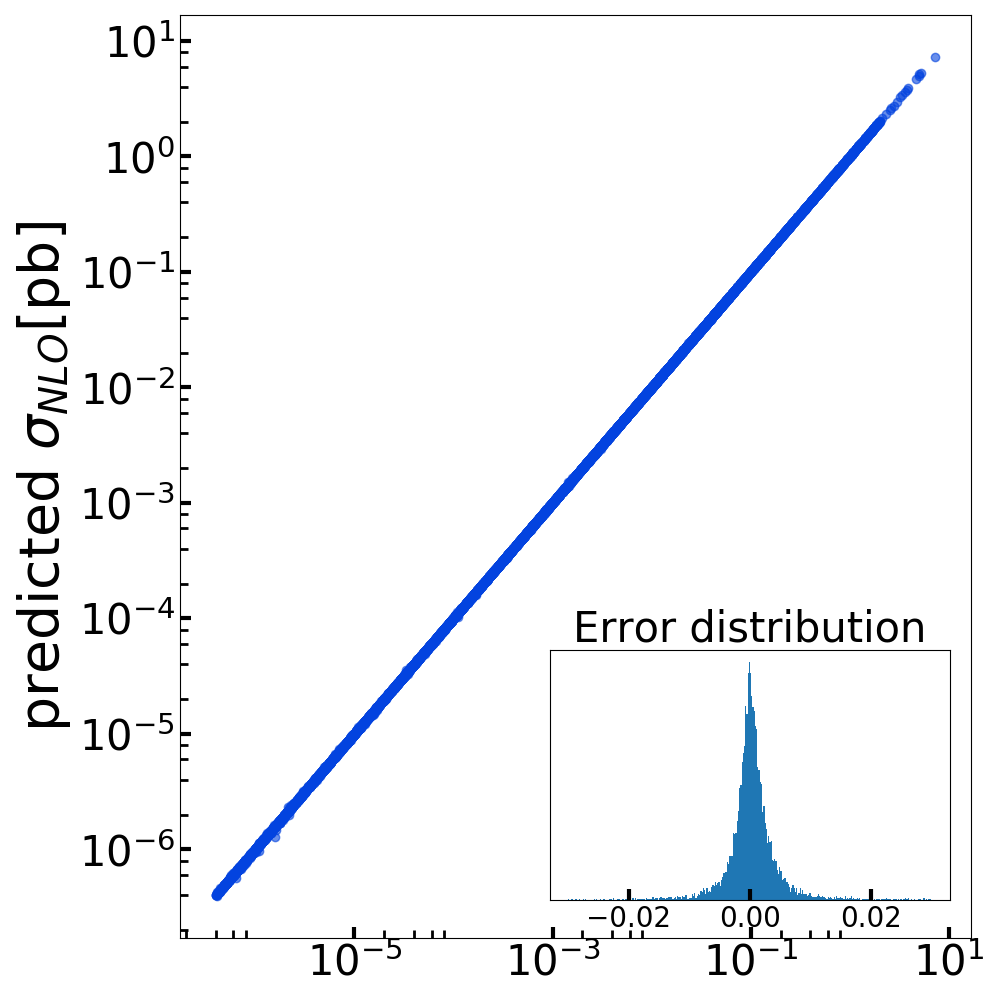}
\includegraphics[width=\textwidth]{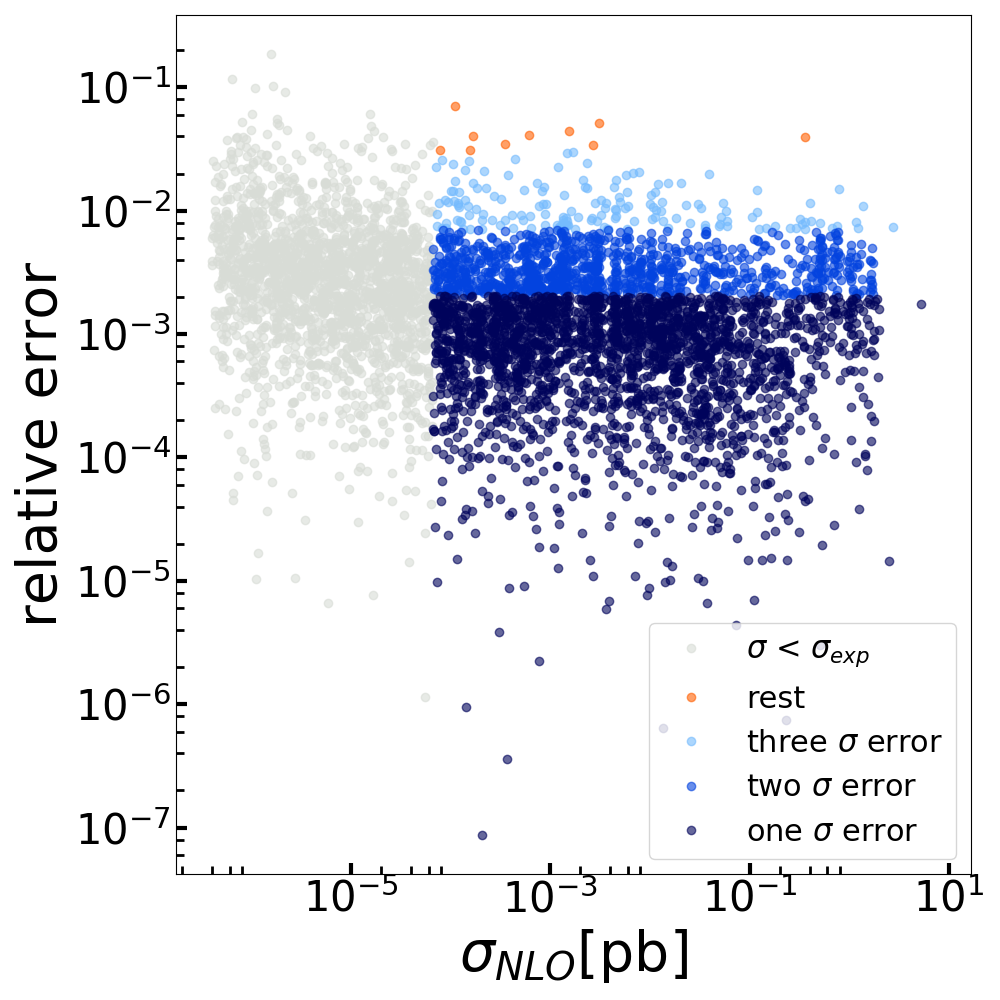}
\subcaption{$\tilde{\chi}^0_2\tilde{\chi}^-_1$}
\end{subfigure}
\caption{The true vs.\ predicted NLO cross sections with histograms of the relative error (top) and the true NLO cross section vs.\ the relative error with confidence intervals (bottom), as defined in Section~\ref{sec:results}, for the test sample of $10^4$ points.\label{fig:error}}
\end{figure*}

\begin{figure*}
\begin{subfigure}{0.49\textwidth}
\includegraphics[width=\textwidth]{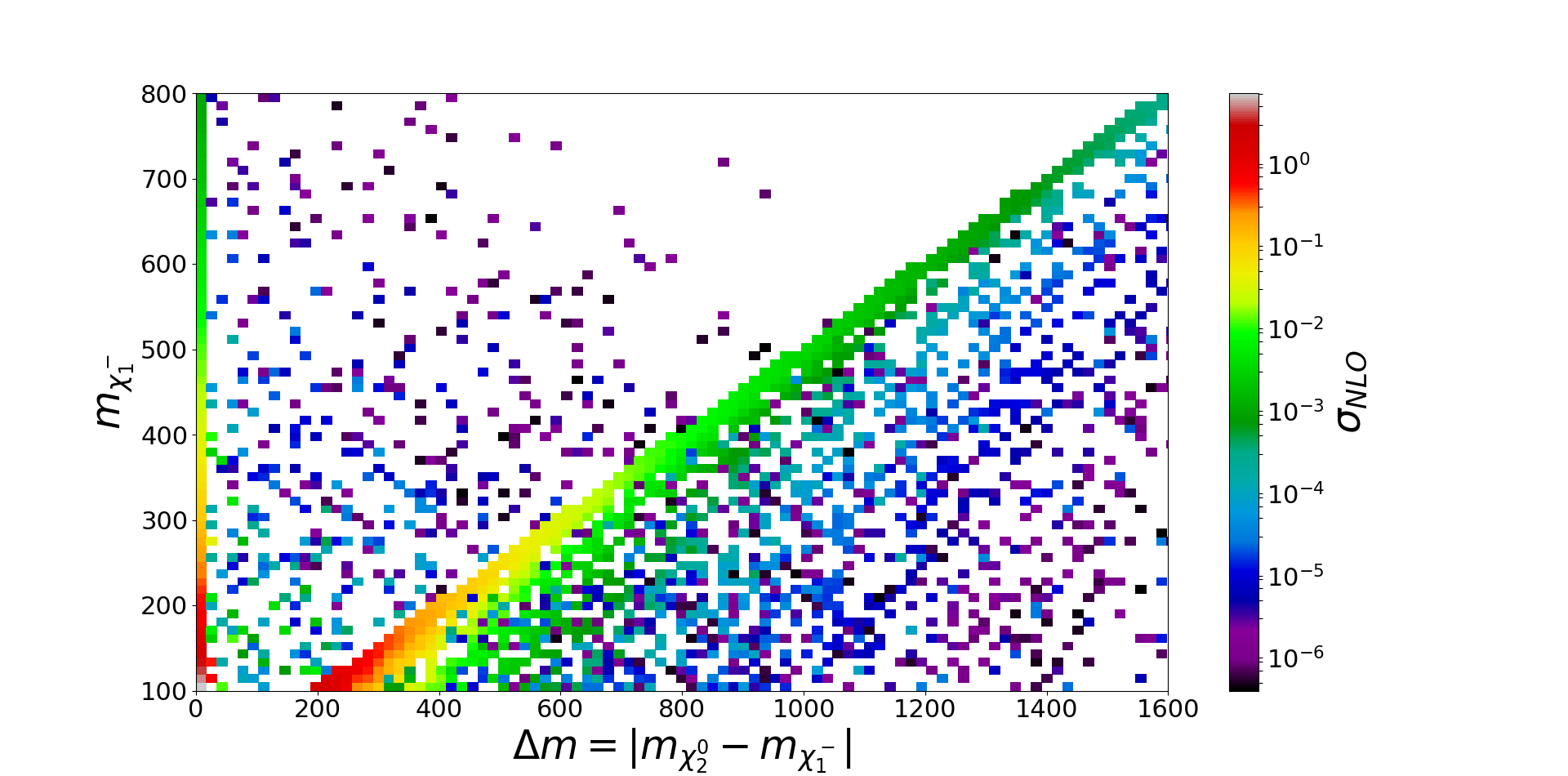}
\includegraphics[width=\textwidth]{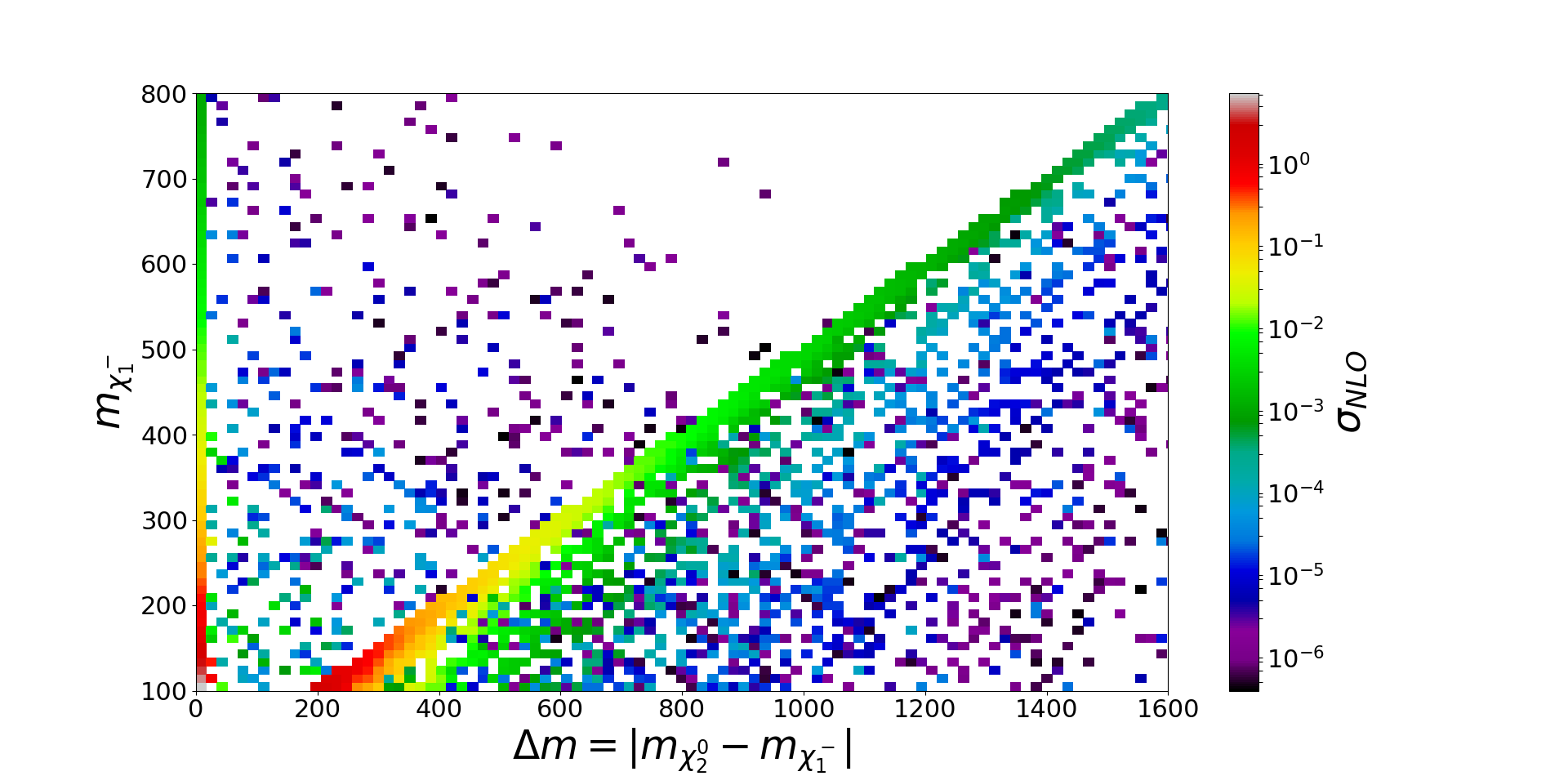}
\includegraphics[width=\textwidth]{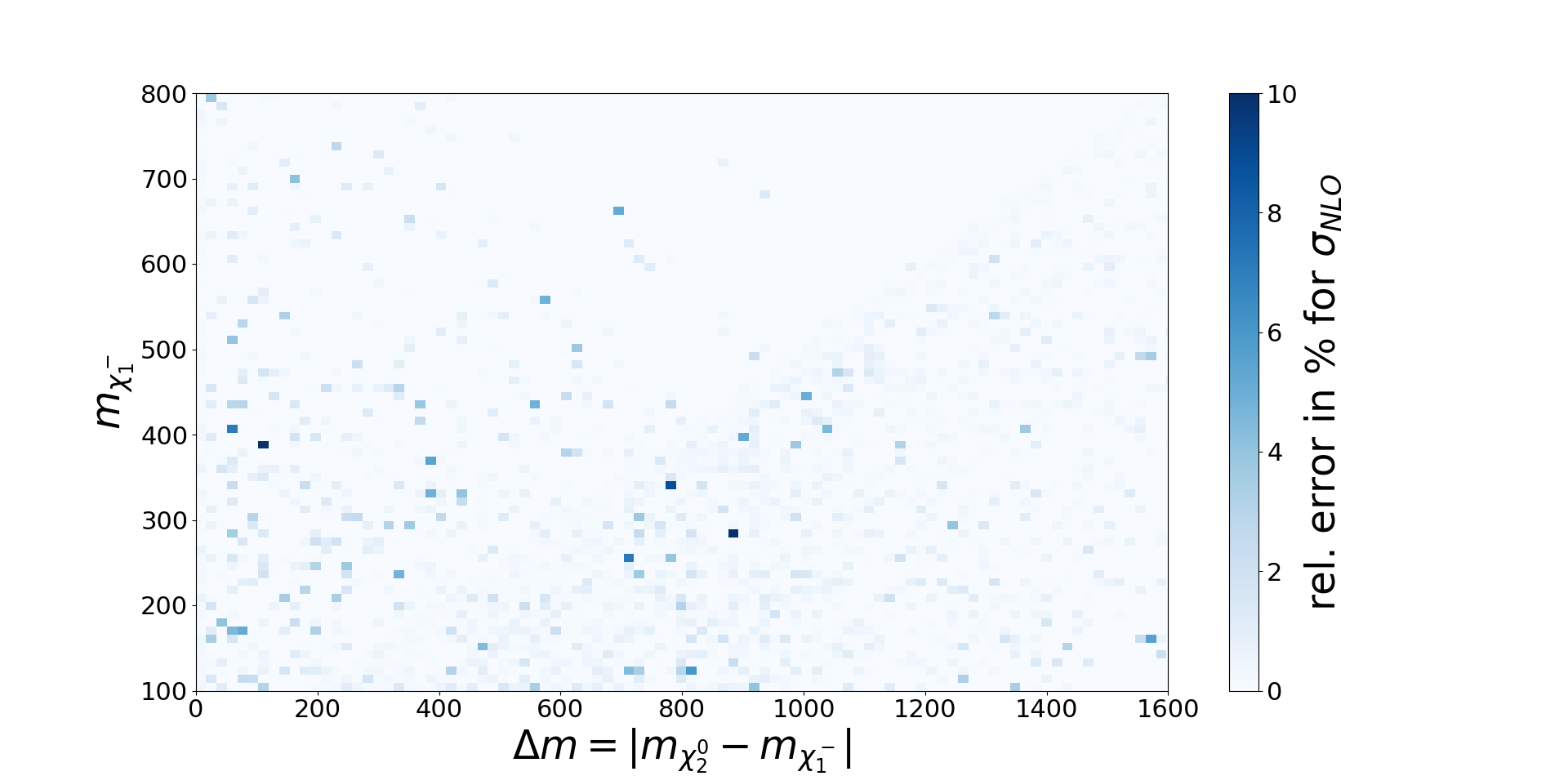}
\subcaption{$\sigma_{\rm{NLO}}(\Delta m,m_{\tilde{\chi}^-_1})$}
\end{subfigure}
\begin{subfigure}{0.49\textwidth}
\includegraphics[width=\textwidth]{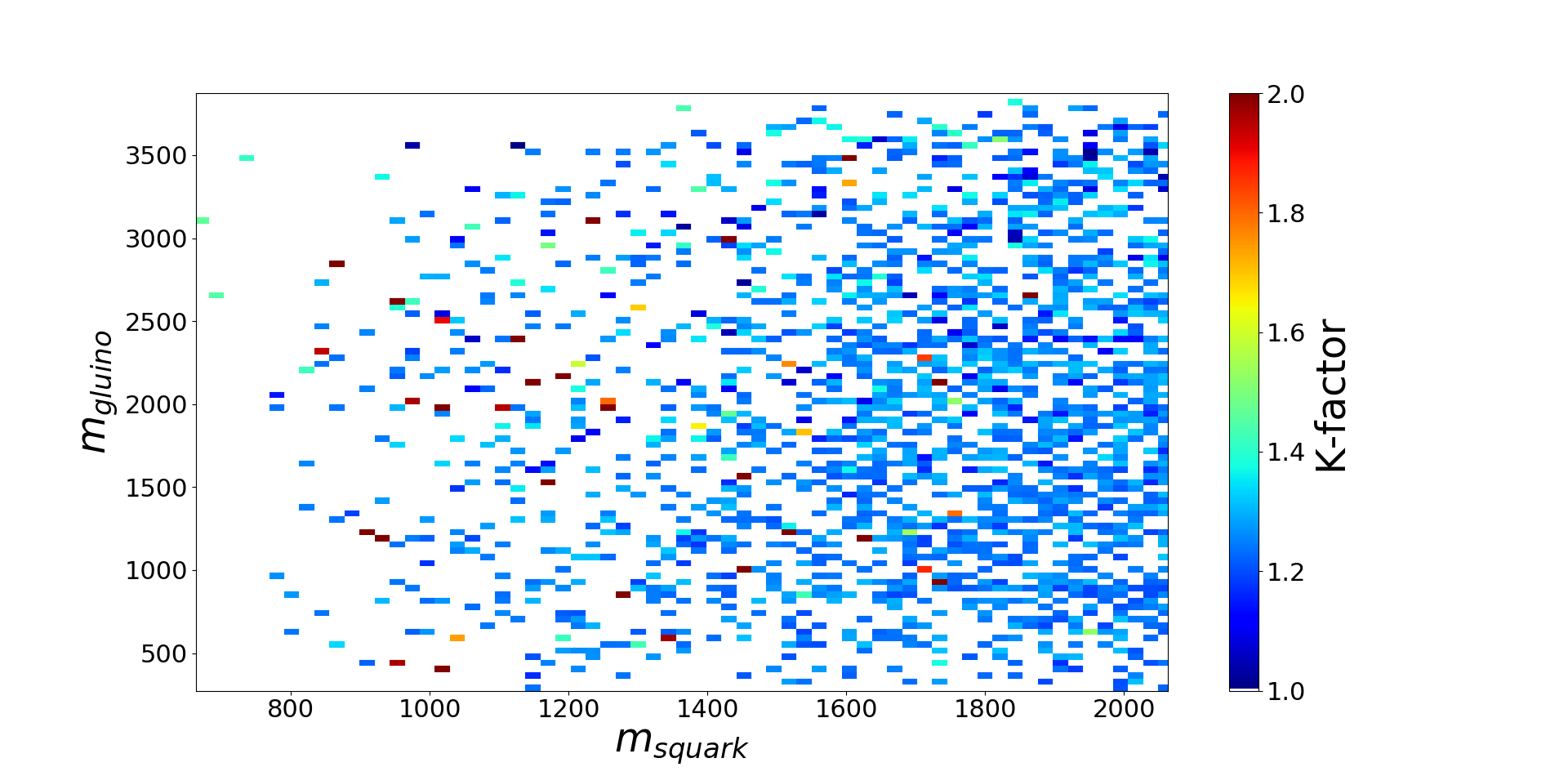}
\includegraphics[width=\textwidth]{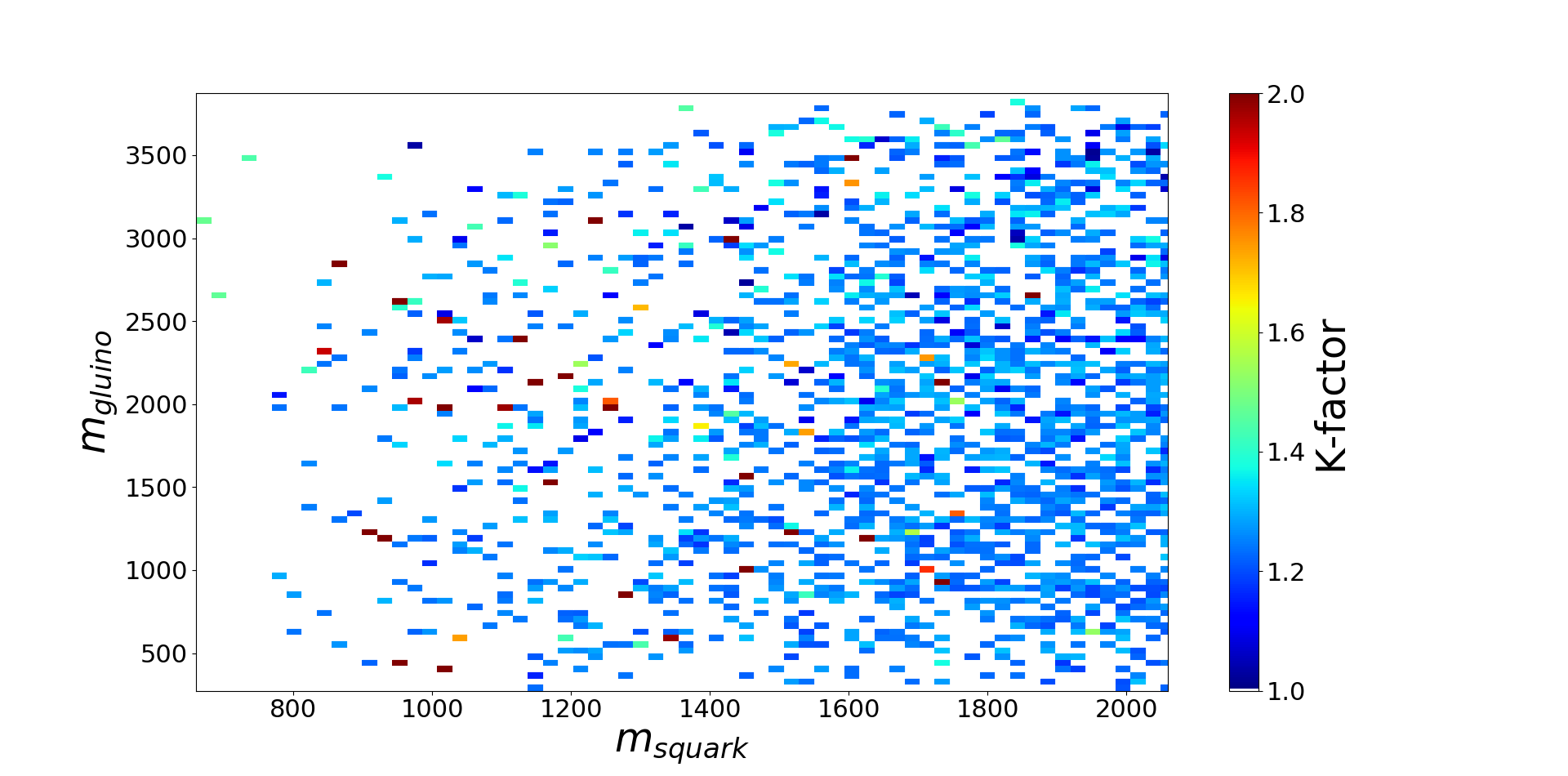}
\includegraphics[width=\textwidth]{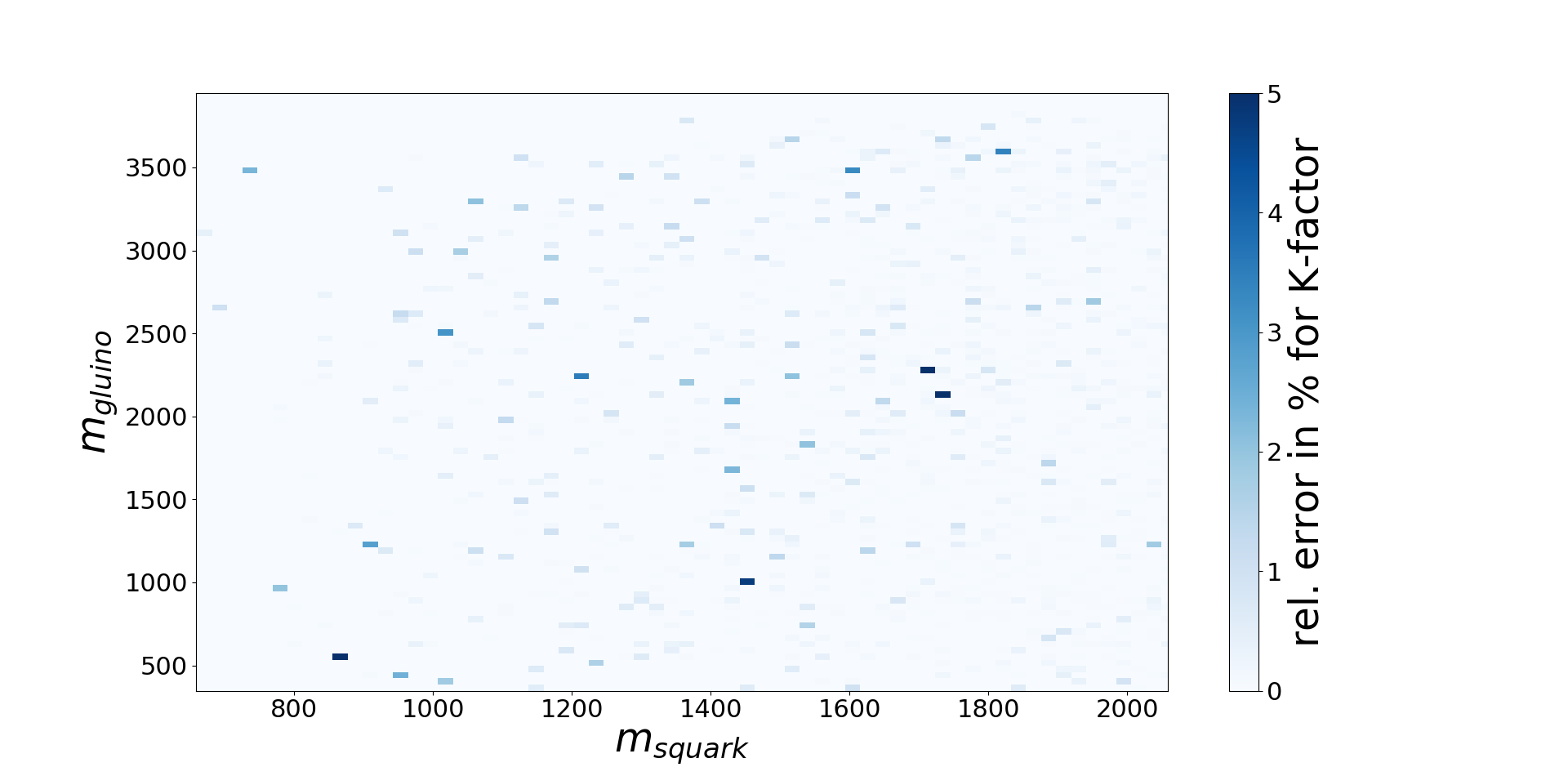}
\subcaption{$K(m_{\tilde{q}},m_{\tilde{g}})$}
\end{subfigure}
\caption{\texttt{Prospino} (top), \texttt{DeepXS} (middle) and relative error (bottom) (a) for $\sigma_{\rm{NLO}}$ in the $\Delta m = |m_{\tilde{\chi}^0_2}-m_{\tilde{\chi}^-_1}|$ vs. $m_{\tilde{\chi}^-_1}$ plane; (b) for the $K$-factor in the $m_{\tilde{q}}=\frac{1}{8}\sum_{i=1}^8 m_{\tilde{q}_i}$ vs. $m_{\tilde{g}}$ plane for $pp\to\tilde{\chi}^0_2\tilde{\chi}^-_1$.}
\label{fig:prospinovsnn}
\end{figure*}

For $\tilde{\chi}^0_2\tilde{\chi}^\pm_1$ we extend our setup to include NNPS because of a recurring problem of outliers with large errors in problematic, often underpopulated, regions of the parameter space even with $10^7$ training samples. NNPS allowed us to have a much better performance with only a fraction of the random samples. The NNPS setup was as follows: the initial training begins with $10^6$ ($\tilde{\chi}^0_2\tilde{\chi}^+_1$) or $1.5\cdot 10^6$ ($\tilde{\chi}^0_2\tilde{\chi}^-_1)$ samples and runs for a short amount of time, namely 40 epochs, 5 iterations and a batch-size of 1000. The resulting neural network is then evaluated on the pool of the remaining $9\cdot 10^6$ samples. The $10^5$ or $1.5\cdot 10^5$ samples for which the neural network performed worst are iteratively added to the training set. This is done 10 and 15 times respectively. The actively sampled training set is then used for a more thorough training identical to the procedure used for $\tilde{\chi}^+_1\tilde{\chi}^-_1$ and $\tilde{\chi}^0_2\tilde{\chi}^0_2$. The evaluation of these networks is then investigated and in both cases the performance is further enhanced by training additional neural networks that are specialised on a fraction of the target value range. For a specialised network covering target values in the range $0.001$ and $0.2$\,pb for $\tilde{\chi}^0_2\tilde{\chi}^-_1$, we also $z$-score transformed the target values without taking the logarithm. The general and specialised networks are then stacked together by using the general network to predict whether a point is predicted better by the specialised network: if that is the case, the prediction of the specialised network is returned, if not, the general network will return its prediction. The $\tilde{\chi}^0_2\tilde{\chi}^+_1$ $K$-factors have been treated similarly, while for the $\tilde{\chi}^0_2\tilde{\chi}^-_1$ we only used one neural network trained on random samples.

\section{\label{sec:results}Results}

In this section we present the accuracy of the tool, \texttt{DeepXS}, including statistical measures of its performance. We also discuss inference times and subtleties of its validity. The testing of \texttt{DeepXS} was performed using $10^4$ pMSSM-19 points generated according to the same rules as the training samples.    

Table~\ref{tab:errorbands} shows the performance of \texttt{DeepXS} for the cross sections $\sigma_{\rm{NLO}}$ that are larger than the threshold $\sigma_{\rm{exp}}=6.6\cdot 10^{-5}\,\rm{pb}$. This threshold corresponds to the integrated luminosity $150\,\rm{fb}^{-1}$, which is the data collected thus far by the LHC, and assuming 10 produced events. Therefore, for current applications this threshold provides a very conservative estimate of the observable electroweakino production. The entries for $1\sigma$, $2\sigma$ and $3\sigma$ denote the maximum error for $68.27\,\%$, $95.45\,\%$ and $99.73\,\%$ of the samples. We use the intervals as defined for the normal distribution motivated by the shape of the error distribution. However, we note that it has fatter tails.

\begin{table}[h]
\begin{center}
\begin{tabular}{|M{1.25cm} | M{1.5cm} | M{1.5cm} | M{1.5cm} | M{1.5cm} |}
\hline
  \textbf{Pair} & \textbf{MAPE} & \textbf{1$\sigma$} & \textbf{2$\sigma$} & \textbf{3$\sigma$} \\
	\thickhline
    $\tilde{\chi}_1^+\tilde{\chi}_1^-$ & 0.091\,\% & 0.081\,\% & 0.334\,\% & 1.195\,\% \\ \hline
    $\tilde{\chi}_2^0\tilde{\chi}_2^0$ & 0.384\,\% & 0.274\,\% & 1.652\,\% & 5.773\,\% \\ \hline
	$\tilde{\chi}_2^0\tilde{\chi}_1^+$ & 0.263\,\% & 0.258\,\% & 0.822\,\% & 3.299\,\% \\ \hline
	$\tilde{\chi}_2^0\tilde{\chi}_1^-$ & 0.214\,\% & 0.206\,\% & 0.701\,\% & 3.035\,\% \\ \hline	
\end{tabular}
\caption{Relative error bands and MAPE for the NLO predictions for $\sigma_{\rm{NLO}}\geq \sigma_{\rm{exp}}=6.6\cdot 10^{-5}$\,pb.}
\label{tab:errorbands}
\end{center}
\end{table}

With all MAPEs being well below $0.5\,\%$ and a maximum error of the $3\sigma$ bands of $5.773\,\%$, the error of the cross sections clearly is sub-dominant relative to scale and PDF uncertainty for a large majority of the presented cases. Figure~\ref{fig:error} demonstrates a large density of points around an error of $10^{-4}$--$10^{-2}$, which matches the precision of Vegas integration, $5\cdot 10^{-3}$, typically reported by \verb+Prospino+. For $\tilde{\chi}^+_1\tilde{\chi}^-_1$, the maximum error on  the $10^4$ test samples is $\approx 3\,\%$ while for the other pairs it is $\mathcal{O}(10\,\%)$. We note that this size of uncertainty is otherwise expected to arise due to PDF and scale variation which starts with $3$--$4\,\%$ for high cross sections and rises to $\mathcal{O}(10\,\%)$ for high masses. The largest error is observed for two samples with an error of $\approx 10\%$ for $\tilde{\chi}^0_2\tilde{\chi}^0_2$ and $\tilde{\chi}^0_2\tilde{\chi}^+_1$. 

The case of $\tilde{\chi}^0_2\tilde{\chi}^0_2$ will be improved with NNPS in the upcoming version of the tool. Note that the dimensionality of $\tilde{\chi}^0_2\tilde{\chi}^\pm_1$, $d=11$ at LO and $d=12$ at NLO, is much higher than for $\tilde{\chi}^0_2\tilde{\chi}^0_2$, with dimensionality 6 and 7 respectively. When training $\tilde{\chi}^0_2\tilde{\chi}^\pm_1$ on $10^7$ random samples, the predictions were much worse than they are currently for $\tilde{\chi}^0_2\tilde{\chi}^0_2$. We thus expect that NNPS will further improve the $\tilde{\chi}^0_2\tilde{\chi}^0_2$ to the same level of precision we have achieved for the other pairs. That the overall accuracy for $\tilde{\chi}^0_2\tilde{\chi}^-_1$ is better than for $\tilde{\chi}^0_2\tilde{\chi}^+_1$ is due to the more thoroughly performed NNPS, which will thus be the standard for future work. 

Below the threshold of $\sigma_{\rm{exp}}$, our predictions also have a MAPE of below $1\%$ with a maximum of $0.81\,\%$ for $\tilde{\chi}^0_2\tilde{\chi}^0_2$, lower values of $\approx 0.3\,\%$ for the mixed pairs and $\approx 0.1\,\%$ for chargino pairs. Note however, that although errors above $10\,\%$ are more frequent for $\sigma\leq\sigma_{\rm{exp}}$, the PDF uncertainty is typically also high in the corresponding region of the parameter space. 

Figure~\ref{fig:prospinovsnn} shows a comparison between the \verb+Prospino+ calculation and the \verb+DeepXS+ prediction including the relative errors for $\sigma_{\rm{NLO}}$ and the $K$-factor for $pp\to\tilde{\chi}^0_2\tilde{\chi}^-_1$. We observe that the neural networks predict the complicated cross section landscapes so well that the plots corresponding to the predictions and the \verb+Prospino+ calculations are indistinguishable by eye. Only the plot showing the relative errors reveal a handful of slight deviations of no more than $\mathcal{O}(10\%)$, consistent with Fig.~\ref{fig:error}. The plots were created using the same 10000 points as for Fig.~\ref{fig:error}. 

\texttt{DeepXS} is interfaced to \texttt{pySLHA}~\cite{Buckley:2013jua} and can process SLHA2 files~\cite{Allanach:2008qq}. Additionally, a possibility to feed in the relevant parameters via .csv and .txt files has been implemented. When providing SLHA files, \texttt{DeepXS} needed $72.3$\,s to evaluate $10^4$ samples or $7.23$ ms per evaluation of $\tilde{\chi}^+_1\tilde{\chi}^-_1$ at LO and NLO, already making \texttt{DeepXS} $\mathcal{O}(10^4)$ faster than \verb+Prospino+. When SLHA files are an input, \texttt{DeepXS} tests if $\tilde{\chi}^0_1$ is the LSP, if the light chargino mass is above $100$\,GeV and if the squark masses are above $500$\,GeV, and a warning is given if any of these conditions is not fulfilled. When text files with an array are provided, the inference of $10^7$ $\tilde{\chi}^+_1\tilde{\chi}^-_1$ predictions both at LO and NLO took 261.51 seconds on an Intel i7-4790K CPU or $\approx 26\,\mu$s per evaluation, making it $\approx 6.9$ million times faster than \verb+Prospino+. When predicting mixed pairs, each evaluation takes slightly longer due to the stacking and the necessity to infer from more than two neural networks. In all cases, warnings are given when the predicted cross section is lower than $\sigma_{\rm{exp}}$.

\section{\label{sec:outlook}Conclusions}

We presented a method that for the first time allows a fast and highly accurate approximation of cross sections that depend on a high-dimensional and complex parameter space. As the first application, we developed a novel tool, \texttt{DeepXS}, that enables a fast approximation of NLO cross sections for pMSSM-19 electroweakinos. Beside the incorporation of expert knowledge, it employs stacked artificial neural networks supplemented by ANN-based point selection techniques to provide fast predictions based on the full NLO calculation using \texttt{Prospino}. Compared to \texttt{Prospino}, \texttt{DeepXS} is more than 4 and up to 7 orders of magnitude faster, while ensuring an accuracy of 1\,\% for more than 95\,\% of the test points. The tool can be found in a GitHub repository~\cite{DeepXS} including examples that show the NNPS sampling strategy. Further development will include the completion of all electroweakino pairs, extensions of the MSSM, an estimation of scale and PDF uncertainties and a merge with \texttt{BSM-AI} \cite{Caron:2016hib}.

\section{Acknowledgements}

S.\ C.\ and S.\ O.\ thank the support by the Netherlands eScience Center under the project iDark: The intelligent Dark Matter Survey. R.\ RdA,  thanks the support from the European Union’s Horizon 2020 research and innovation programme under the Marie Sk\l{}odowska-Curie grant agreement No 674896, the ``SOM Sabor y origen de la Materia" MEC projects and the Spanish MINECO Centro de Excelencia Severo Ochoa del IFIC program under grant SEV-2014-0398. K.\ R.\ is supported by the National Science Centre, Poland, under research grants DEC-2016/23/G/ST2/04301, 2015/18/M/ST2/00518 and 2015/19/D/ST2/03136.

\bibliographystyle{apsrev4-1}
\bibliography{bibtex}

\begin{thebibliography}{34}%
\makeatletter
\providecommand \@ifxundefined [1]{%
 \@ifx{#1\undefined}
}%
\providecommand \@ifnum [1]{%
 \ifnum #1\expandafter \@firstoftwo
 \else \expandafter \@secondoftwo
 \fi
}%
\providecommand \@ifx [1]{%
 \ifx #1\expandafter \@firstoftwo
 \else \expandafter \@secondoftwo
 \fi
}%
\providecommand \natexlab [1]{#1}%
\providecommand \enquote  [1]{``#1''}%
\providecommand \bibnamefont  [1]{#1}%
\providecommand \bibfnamefont [1]{#1}%
\providecommand \citenamefont [1]{#1}%
\providecommand \href@noop [0]{\@secondoftwo}%
\providecommand \href [0]{\begingroup \@sanitize@url \@href}%
\providecommand \@href[1]{\@@startlink{#1}\@@href}%
\providecommand \@@href[1]{\endgroup#1\@@endlink}%
\providecommand \@sanitize@url [0]{\catcode `\\12\catcode `\$12\catcode
  `\&12\catcode `\#12\catcode `\^12\catcode `\_12\catcode `\%12\relax}%
\providecommand \@@startlink[1]{}%
\providecommand \@@endlink[0]{}%
\providecommand \url  [0]{\begingroup\@sanitize@url \@url }%
\providecommand \@url [1]{\endgroup\@href {#1}{\urlprefix }}%
\providecommand \urlprefix  [0]{URL }%
\providecommand \Eprint [0]{\href }%
\providecommand \doibase [0]{http://dx.doi.org/}%
\providecommand \selectlanguage [0]{\@gobble}%
\providecommand \bibinfo  [0]{\@secondoftwo}%
\providecommand \bibfield  [0]{\@secondoftwo}%
\providecommand \translation [1]{[#1]}%
\providecommand \BibitemOpen [0]{}%
\providecommand \bibitemStop [0]{}%
\providecommand \bibitemNoStop [0]{.\EOS\space}%
\providecommand \EOS [0]{\spacefactor3000\relax}%
\providecommand \BibitemShut  [1]{\csname bibitem#1\endcsname}%
\let\auto@bib@innerbib\@empty
\bibitem [{\citenamefont {Abdallah}\ \emph {et~al.}(2015)\citenamefont
  {Abdallah} \emph {et~al.}}]{Abdallah:2015ter}%
  \BibitemOpen
  \bibfield  {author} {\bibinfo {author} {\bibfnamefont {J.}~\bibnamefont
  {Abdallah}} \emph {et~al.},\ }\href {\doibase 10.1016/j.dark.2015.08.001}
  {\bibfield  {journal} {\bibinfo  {journal} {Phys. Dark Univ.}\ }\textbf
  {\bibinfo {volume} {9-10}},\ \bibinfo {pages} {8} (\bibinfo {year} {2015})},\
  \Eprint {http://arxiv.org/abs/1506.03116} {arXiv:1506.03116 [hep-ph]}
  \BibitemShut {NoStop}%
\bibitem [{\citenamefont {Athron}\ \emph {et~al.}(2017)\citenamefont {Athron}
  \emph {et~al.}}]{Athron:2017yua}%
  \BibitemOpen
  \bibfield  {author} {\bibinfo {author} {\bibfnamefont {P.}~\bibnamefont
  {Athron}} \emph {et~al.} (\bibinfo {collaboration} {GAMBIT}),\ }\href
  {\doibase 10.1140/epjc/s10052-017-5196-8} {\bibfield  {journal} {\bibinfo
  {journal} {Eur. Phys. J.}\ }\textbf {\bibinfo {volume} {C77}},\ \bibinfo
  {pages} {879} (\bibinfo {year} {2017})},\ \Eprint
  {http://arxiv.org/abs/1705.07917} {arXiv:1705.07917 [hep-ph]} \BibitemShut
  {NoStop}%
\bibitem [{\citenamefont {Bertone}\ \emph {et~al.}(2018)\citenamefont
  {Bertone}, \citenamefont {Bozorgnia}, \citenamefont {Kim}, \citenamefont
  {Liem}, \citenamefont {McCabe}, \citenamefont {Otten},\ and\ \citenamefont
  {Ruiz~de Austri}}]{Bertone:2017adx}%
  \BibitemOpen
  \bibfield  {author} {\bibinfo {author} {\bibfnamefont {G.}~\bibnamefont
  {Bertone}}, \bibinfo {author} {\bibfnamefont {N.}~\bibnamefont {Bozorgnia}},
  \bibinfo {author} {\bibfnamefont {J.~S.}\ \bibnamefont {Kim}}, \bibinfo
  {author} {\bibfnamefont {S.}~\bibnamefont {Liem}}, \bibinfo {author}
  {\bibfnamefont {C.}~\bibnamefont {McCabe}}, \bibinfo {author} {\bibfnamefont
  {S.}~\bibnamefont {Otten}}, \ and\ \bibinfo {author} {\bibfnamefont
  {R.}~\bibnamefont {Ruiz~de Austri}},\ }\href {\doibase
  10.1088/1475-7516/2018/03/026} {\bibfield  {journal} {\bibinfo  {journal}
  {JCAP}\ }\textbf {\bibinfo {volume} {1803}},\ \bibinfo {pages} {026}
  (\bibinfo {year} {2018})},\ \Eprint {http://arxiv.org/abs/1712.04793}
  {arXiv:1712.04793 [hep-ph]} \BibitemShut {NoStop}%
\bibitem [{\citenamefont {Drees}\ \emph {et~al.}(2004)\citenamefont {Drees},
  \citenamefont {Godbole},\ and\ \citenamefont {Roy}}]{Drees:2004jm}%
  \BibitemOpen
  \bibfield  {author} {\bibinfo {author} {\bibfnamefont {M.}~\bibnamefont
  {Drees}}, \bibinfo {author} {\bibfnamefont {R.}~\bibnamefont {Godbole}}, \
  and\ \bibinfo {author} {\bibfnamefont {P.}~\bibnamefont {Roy}},\ }\href@noop
  {} {\emph {\bibinfo {title} {{Theory and phenomenology of sparticles: An
  account of four-dimensional N=1 supersymmetry in high energy physics}}}}\
  (\bibinfo {year} {2004})\BibitemShut {NoStop}%
\bibitem [{\citenamefont {Haber}\ and\ \citenamefont
  {Kane}(1985)}]{Haber:1984rc}%
  \BibitemOpen
  \bibfield  {author} {\bibinfo {author} {\bibfnamefont {H.~E.}\ \bibnamefont
  {Haber}}\ and\ \bibinfo {author} {\bibfnamefont {G.~L.}\ \bibnamefont
  {Kane}},\ }\href {\doibase 10.1016/0370-1573(85)90051-1} {\bibfield
  {journal} {\bibinfo  {journal} {Phys. Rept.}\ }\textbf {\bibinfo {volume}
  {117}},\ \bibinfo {pages} {75} (\bibinfo {year} {1985})}\BibitemShut
  {NoStop}%
\bibitem [{\citenamefont {Martin}(1997)}]{Martin:1997ns}%
  \BibitemOpen
  \bibfield  {author} {\bibinfo {author} {\bibfnamefont {S.~P.}\ \bibnamefont
  {Martin}},\ }\href {\doibase 10.1142/9789812839657_0001,
  10.1142/9789814307505_0001} {\ ,\ \bibinfo {pages} {1} (\bibinfo {year}
  {1997})},\ \bibinfo {note} {[Adv. Ser. Direct. High Energy
  Phys.18,1(1998)]},\ \Eprint {http://arxiv.org/abs/hep-ph/9709356}
  {arXiv:hep-ph/9709356 [hep-ph]} \BibitemShut {NoStop}%
\bibitem [{\citenamefont {Fuks}\ \emph {et~al.}(2012)\citenamefont {Fuks},
  \citenamefont {Klasen}, \citenamefont {Lamprea},\ and\ \citenamefont
  {Rothering}}]{Fuks:2012qx}%
  \BibitemOpen
  \bibfield  {author} {\bibinfo {author} {\bibfnamefont {B.}~\bibnamefont
  {Fuks}}, \bibinfo {author} {\bibfnamefont {M.}~\bibnamefont {Klasen}},
  \bibinfo {author} {\bibfnamefont {D.~R.}\ \bibnamefont {Lamprea}}, \ and\
  \bibinfo {author} {\bibfnamefont {M.}~\bibnamefont {Rothering}},\ }\href
  {\doibase 10.1007/JHEP10(2012)081} {\bibfield  {journal} {\bibinfo  {journal}
  {JHEP}\ }\textbf {\bibinfo {volume} {10}},\ \bibinfo {pages} {081} (\bibinfo
  {year} {2012})},\ \Eprint {http://arxiv.org/abs/1207.2159} {arXiv:1207.2159
  [hep-ph]} \BibitemShut {NoStop}%
\bibitem [{\citenamefont {Beenakker}\ \emph {et~al.}(2014)\citenamefont
  {Beenakker}, \citenamefont {Borschensky}, \citenamefont {Kraemer},
  \citenamefont {Kulesza}, \citenamefont {Laenen}, \citenamefont {Theeuwes},\
  and\ \citenamefont {Thewes}}]{Beenakker:2014sma}%
  \BibitemOpen
  \bibfield  {author} {\bibinfo {author} {\bibfnamefont {W.}~\bibnamefont
  {Beenakker}}, \bibinfo {author} {\bibfnamefont {C.}~\bibnamefont
  {Borschensky}}, \bibinfo {author} {\bibfnamefont {M.}~\bibnamefont
  {Kraemer}}, \bibinfo {author} {\bibfnamefont {A.}~\bibnamefont {Kulesza}},
  \bibinfo {author} {\bibfnamefont {E.}~\bibnamefont {Laenen}}, \bibinfo
  {author} {\bibfnamefont {V.}~\bibnamefont {Theeuwes}}, \ and\ \bibinfo
  {author} {\bibfnamefont {S.}~\bibnamefont {Thewes}},\ }\href {\doibase
  10.1007/JHEP12(2014)023} {\bibfield  {journal} {\bibinfo  {journal} {JHEP}\
  }\textbf {\bibinfo {volume} {12}},\ \bibinfo {pages} {023} (\bibinfo {year}
  {2014})},\ \Eprint {http://arxiv.org/abs/1404.3134} {arXiv:1404.3134
  [hep-ph]} \BibitemShut {NoStop}%
\bibitem [{\citenamefont {Beenakker}\ \emph {et~al.}(1999)\citenamefont
  {Beenakker}, \citenamefont {Klasen}, \citenamefont {Kraemer}, \citenamefont
  {Plehn}, \citenamefont {Spira},\ and\ \citenamefont
  {Zerwas}}]{Beenakker:1999xh}%
  \BibitemOpen
  \bibfield  {author} {\bibinfo {author} {\bibfnamefont {W.}~\bibnamefont
  {Beenakker}}, \bibinfo {author} {\bibfnamefont {M.}~\bibnamefont {Klasen}},
  \bibinfo {author} {\bibfnamefont {M.}~\bibnamefont {Kraemer}}, \bibinfo
  {author} {\bibfnamefont {T.}~\bibnamefont {Plehn}}, \bibinfo {author}
  {\bibfnamefont {M.}~\bibnamefont {Spira}}, \ and\ \bibinfo {author}
  {\bibfnamefont {P.~M.}\ \bibnamefont {Zerwas}},\ }\href {\doibase
  10.1103/PhysRevLett.100.029901, 10.1103/PhysRevLett.83.3780} {\bibfield
  {journal} {\bibinfo  {journal} {Phys. Rev. Lett.}\ }\textbf {\bibinfo
  {volume} {83}},\ \bibinfo {pages} {3780} (\bibinfo {year} {1999})},\ \bibinfo
  {note} {[Erratum: Phys. Rev. Lett.100,029901(2008)]},\ \Eprint
  {http://arxiv.org/abs/hep-ph/9906298} {arXiv:hep-ph/9906298 [hep-ph]}
  \BibitemShut {NoStop}%
\bibitem [{\citenamefont {Fuks}\ \emph {et~al.}(2013)\citenamefont {Fuks},
  \citenamefont {Klasen}, \citenamefont {Lamprea},\ and\ \citenamefont
  {Rothering}}]{Fuks:2013vua}%
  \BibitemOpen
  \bibfield  {author} {\bibinfo {author} {\bibfnamefont {B.}~\bibnamefont
  {Fuks}}, \bibinfo {author} {\bibfnamefont {M.}~\bibnamefont {Klasen}},
  \bibinfo {author} {\bibfnamefont {D.~R.}\ \bibnamefont {Lamprea}}, \ and\
  \bibinfo {author} {\bibfnamefont {M.}~\bibnamefont {Rothering}},\ }\href
  {\doibase 10.1140/epjc/s10052-013-2480-0} {\bibfield  {journal} {\bibinfo
  {journal} {Eur. Phys. J.}\ }\textbf {\bibinfo {volume} {C73}},\ \bibinfo
  {pages} {2480} (\bibinfo {year} {2013})},\ \Eprint
  {http://arxiv.org/abs/1304.0790} {arXiv:1304.0790 [hep-ph]} \BibitemShut
  {NoStop}%
\bibitem [{\citenamefont {Bechtle}\ \emph {et~al.}(2012)\citenamefont {Bechtle}
  \emph {et~al.}}]{Bechtle:2012zk}%
  \BibitemOpen
  \bibfield  {author} {\bibinfo {author} {\bibfnamefont {P.}~\bibnamefont
  {Bechtle}} \emph {et~al.},\ }\href {\doibase 10.1007/JHEP06(2012)098}
  {\bibfield  {journal} {\bibinfo  {journal} {JHEP}\ }\textbf {\bibinfo
  {volume} {06}},\ \bibinfo {pages} {098} (\bibinfo {year} {2012})},\ \Eprint
  {http://arxiv.org/abs/1204.4199} {arXiv:1204.4199 [hep-ph]} \BibitemShut
  {NoStop}%
\bibitem [{\citenamefont {Strege}\ \emph {et~al.}(2014)\citenamefont {Strege},
  \citenamefont {Bertone}, \citenamefont {Besjes}, \citenamefont {Caron},
  \citenamefont {Ruiz~de Austri}, \citenamefont {Strubig},\ and\ \citenamefont
  {Trotta}}]{Strege:2014ija}%
  \BibitemOpen
  \bibfield  {author} {\bibinfo {author} {\bibfnamefont {C.}~\bibnamefont
  {Strege}}, \bibinfo {author} {\bibfnamefont {G.}~\bibnamefont {Bertone}},
  \bibinfo {author} {\bibfnamefont {G.~J.}\ \bibnamefont {Besjes}}, \bibinfo
  {author} {\bibfnamefont {S.}~\bibnamefont {Caron}}, \bibinfo {author}
  {\bibfnamefont {R.}~\bibnamefont {Ruiz~de Austri}}, \bibinfo {author}
  {\bibfnamefont {A.}~\bibnamefont {Strubig}}, \ and\ \bibinfo {author}
  {\bibfnamefont {R.}~\bibnamefont {Trotta}},\ }\href {\doibase
  10.1007/JHEP09(2014)081} {\bibfield  {journal} {\bibinfo  {journal} {JHEP}\
  }\textbf {\bibinfo {volume} {09}},\ \bibinfo {pages} {081} (\bibinfo {year}
  {2014})},\ \Eprint {http://arxiv.org/abs/1405.0622} {arXiv:1405.0622
  [hep-ph]} \BibitemShut {NoStop}%
\bibitem [{\citenamefont {Bagnaschi}\ \emph {et~al.}(2018)\citenamefont
  {Bagnaschi} \emph {et~al.}}]{Bagnaschi:2017tru}%
  \BibitemOpen
  \bibfield  {author} {\bibinfo {author} {\bibfnamefont {E.}~\bibnamefont
  {Bagnaschi}} \emph {et~al.},\ }\href {\doibase
  10.1140/epjc/s10052-018-5697-0} {\bibfield  {journal} {\bibinfo  {journal}
  {Eur. Phys. J.}\ }\textbf {\bibinfo {volume} {C78}},\ \bibinfo {pages} {256}
  (\bibinfo {year} {2018})},\ \Eprint {http://arxiv.org/abs/1710.11091}
  {arXiv:1710.11091 [hep-ph]} \BibitemShut {NoStop}%
\bibitem [{\citenamefont {Athron}\ \emph {et~al.}(2018)\citenamefont {Athron}
  \emph {et~al.}}]{Athron:2018vxy}%
  \BibitemOpen
  \bibfield  {author} {\bibinfo {author} {\bibfnamefont {P.}~\bibnamefont
  {Athron}} \emph {et~al.} (\bibinfo {collaboration} {GAMBIT}),\ }\href@noop {}
  {\  (\bibinfo {year} {2018})},\ \Eprint {http://arxiv.org/abs/1809.02097}
  {arXiv:1809.02097 [hep-ph]} \BibitemShut {NoStop}%
\bibitem [{\citenamefont {Bertone}\ \emph {et~al.}(2016)\citenamefont
  {Bertone}, \citenamefont {Calore}, \citenamefont {Caron}, \citenamefont
  {Ruiz}, \citenamefont {Kim}, \citenamefont {Trotta},\ and\ \citenamefont
  {Weniger}}]{Bertone:2015tza}%
  \BibitemOpen
  \bibfield  {author} {\bibinfo {author} {\bibfnamefont {G.}~\bibnamefont
  {Bertone}}, \bibinfo {author} {\bibfnamefont {F.}~\bibnamefont {Calore}},
  \bibinfo {author} {\bibfnamefont {S.}~\bibnamefont {Caron}}, \bibinfo
  {author} {\bibfnamefont {R.}~\bibnamefont {Ruiz}}, \bibinfo {author}
  {\bibfnamefont {J.~S.}\ \bibnamefont {Kim}}, \bibinfo {author} {\bibfnamefont
  {R.}~\bibnamefont {Trotta}}, \ and\ \bibinfo {author} {\bibfnamefont
  {C.}~\bibnamefont {Weniger}},\ }\href {\doibase
  10.1088/1475-7516/2016/04/037} {\bibfield  {journal} {\bibinfo  {journal}
  {JCAP}\ }\textbf {\bibinfo {volume} {1604}},\ \bibinfo {pages} {037}
  (\bibinfo {year} {2016})},\ \Eprint {http://arxiv.org/abs/1507.07008}
  {arXiv:1507.07008 [hep-ph]} \BibitemShut {NoStop}%
\bibitem [{\citenamefont {Kim}\ \emph {et~al.}(2016)\citenamefont {Kim},
  \citenamefont {Rolbiecki}, \citenamefont {Ruiz}, \citenamefont {Tattersall},\
  and\ \citenamefont {Weber}}]{Kim:2016rsd}%
  \BibitemOpen
  \bibfield  {author} {\bibinfo {author} {\bibfnamefont {J.~S.}\ \bibnamefont
  {Kim}}, \bibinfo {author} {\bibfnamefont {K.}~\bibnamefont {Rolbiecki}},
  \bibinfo {author} {\bibfnamefont {R.}~\bibnamefont {Ruiz}}, \bibinfo {author}
  {\bibfnamefont {J.}~\bibnamefont {Tattersall}}, \ and\ \bibinfo {author}
  {\bibfnamefont {T.}~\bibnamefont {Weber}},\ }\href {\doibase
  10.1103/PhysRevD.94.095013} {\bibfield  {journal} {\bibinfo  {journal} {Phys.
  Rev.}\ }\textbf {\bibinfo {volume} {D94}},\ \bibinfo {pages} {095013}
  (\bibinfo {year} {2016})},\ \Eprint {http://arxiv.org/abs/1606.06738}
  {arXiv:1606.06738 [hep-ph]} \BibitemShut {NoStop}%
\bibitem [{\citenamefont {Beenakker}\ \emph {et~al.}(2016)\citenamefont
  {Beenakker}, \citenamefont {Borschensky}, \citenamefont {Kr{\"{a}}mer},
  \citenamefont {Kulesza},\ and\ \citenamefont {Laenen}}]{Beenakker:2016lwe}%
  \BibitemOpen
  \bibfield  {author} {\bibinfo {author} {\bibfnamefont {W.}~\bibnamefont
  {Beenakker}}, \bibinfo {author} {\bibfnamefont {C.}~\bibnamefont
  {Borschensky}}, \bibinfo {author} {\bibfnamefont {M.}~\bibnamefont
  {Kr{\"{a}}mer}}, \bibinfo {author} {\bibfnamefont {A.}~\bibnamefont
  {Kulesza}}, \ and\ \bibinfo {author} {\bibfnamefont {E.}~\bibnamefont
  {Laenen}},\ }\href {\doibase 10.1007/JHEP12(2016)133} {\bibfield  {journal}
  {\bibinfo  {journal} {JHEP}\ }\textbf {\bibinfo {volume} {12}},\ \bibinfo
  {pages} {133} (\bibinfo {year} {2016})},\ \Eprint
  {http://arxiv.org/abs/1607.07741} {arXiv:1607.07741 [hep-ph]} \BibitemShut
  {NoStop}%
\bibitem [{\citenamefont {Ellwanger}\ \emph {et~al.}(2010)\citenamefont
  {Ellwanger}, \citenamefont {Hugonie},\ and\ \citenamefont
  {Teixeira}}]{Ellwanger:2009dp}%
  \BibitemOpen
  \bibfield  {author} {\bibinfo {author} {\bibfnamefont {U.}~\bibnamefont
  {Ellwanger}}, \bibinfo {author} {\bibfnamefont {C.}~\bibnamefont {Hugonie}},
  \ and\ \bibinfo {author} {\bibfnamefont {A.~M.}\ \bibnamefont {Teixeira}},\
  }\href {\doibase 10.1016/j.physrep.2010.07.001} {\bibfield  {journal}
  {\bibinfo  {journal} {Phys. Rept.}\ }\textbf {\bibinfo {volume} {496}},\
  \bibinfo {pages} {1} (\bibinfo {year} {2010})},\ \Eprint
  {http://arxiv.org/abs/0910.1785} {arXiv:0910.1785 [hep-ph]} \BibitemShut
  {NoStop}%
\bibitem [{\citenamefont {Aad}\ \emph {et~al.}(2015)\citenamefont {Aad} \emph
  {et~al.}}]{Aad:2015baa}%
  \BibitemOpen
  \bibfield  {author} {\bibinfo {author} {\bibfnamefont {G.}~\bibnamefont
  {Aad}} \emph {et~al.} (\bibinfo {collaboration} {ATLAS}),\ }\href {\doibase
  10.1007/JHEP10(2015)134} {\bibfield  {journal} {\bibinfo  {journal} {JHEP}\
  }\textbf {\bibinfo {volume} {10}},\ \bibinfo {pages} {134} (\bibinfo {year}
  {2015})},\ \Eprint {http://arxiv.org/abs/1508.06608} {arXiv:1508.06608
  [hep-ex]} \BibitemShut {NoStop}%
\bibitem [{\citenamefont {Porod}(2003)}]{Porod:2003um}%
  \BibitemOpen
  \bibfield  {author} {\bibinfo {author} {\bibfnamefont {W.}~\bibnamefont
  {Porod}},\ }\href {\doibase 10.1016/S0010-4655(03)00222-4} {\bibfield
  {journal} {\bibinfo  {journal} {Comput. Phys. Commun.}\ }\textbf {\bibinfo
  {volume} {153}},\ \bibinfo {pages} {275} (\bibinfo {year} {2003})},\ \Eprint
  {http://arxiv.org/abs/hep-ph/0301101} {arXiv:hep-ph/0301101 [hep-ph]}
  \BibitemShut {NoStop}%
\bibitem [{\citenamefont {Porod}\ and\ \citenamefont
  {Staub}(2012)}]{Porod:2011nf}%
  \BibitemOpen
  \bibfield  {author} {\bibinfo {author} {\bibfnamefont {W.}~\bibnamefont
  {Porod}}\ and\ \bibinfo {author} {\bibfnamefont {F.}~\bibnamefont {Staub}},\
  }\href {\doibase 10.1016/j.cpc.2012.05.021} {\bibfield  {journal} {\bibinfo
  {journal} {Comput. Phys. Commun.}\ }\textbf {\bibinfo {volume} {183}},\
  \bibinfo {pages} {2458} (\bibinfo {year} {2012})},\ \Eprint
  {http://arxiv.org/abs/1104.1573} {arXiv:1104.1573 [hep-ph]} \BibitemShut
  {NoStop}%
\bibitem [{\citenamefont {Pumplin}\ \emph {et~al.}(2002)\citenamefont
  {Pumplin}, \citenamefont {Stump}, \citenamefont {Huston}, \citenamefont
  {Lai}, \citenamefont {Nadolsky},\ and\ \citenamefont
  {Tung}}]{Pumplin:2002vw}%
  \BibitemOpen
  \bibfield  {author} {\bibinfo {author} {\bibfnamefont {J.}~\bibnamefont
  {Pumplin}}, \bibinfo {author} {\bibfnamefont {D.~R.}\ \bibnamefont {Stump}},
  \bibinfo {author} {\bibfnamefont {J.}~\bibnamefont {Huston}}, \bibinfo
  {author} {\bibfnamefont {H.~L.}\ \bibnamefont {Lai}}, \bibinfo {author}
  {\bibfnamefont {P.~M.}\ \bibnamefont {Nadolsky}}, \ and\ \bibinfo {author}
  {\bibfnamefont {W.~K.}\ \bibnamefont {Tung}},\ }\href {\doibase
  10.1088/1126-6708/2002/07/012} {\bibfield  {journal} {\bibinfo  {journal}
  {JHEP}\ }\textbf {\bibinfo {volume} {07}},\ \bibinfo {pages} {012} (\bibinfo
  {year} {2002})},\ \Eprint {http://arxiv.org/abs/hep-ph/0201195}
  {arXiv:hep-ph/0201195 [hep-ph]} \BibitemShut {NoStop}%
\bibitem [{\citenamefont {Nadolsky}\ \emph {et~al.}(2008)\citenamefont
  {Nadolsky}, \citenamefont {Lai}, \citenamefont {Cao}, \citenamefont {Huston},
  \citenamefont {Pumplin}, \citenamefont {Stump}, \citenamefont {Tung},\ and\
  \citenamefont {Yuan}}]{Nadolsky:2008zw}%
  \BibitemOpen
  \bibfield  {author} {\bibinfo {author} {\bibfnamefont {P.~M.}\ \bibnamefont
  {Nadolsky}}, \bibinfo {author} {\bibfnamefont {H.-L.}\ \bibnamefont {Lai}},
  \bibinfo {author} {\bibfnamefont {Q.-H.}\ \bibnamefont {Cao}}, \bibinfo
  {author} {\bibfnamefont {J.}~\bibnamefont {Huston}}, \bibinfo {author}
  {\bibfnamefont {J.}~\bibnamefont {Pumplin}}, \bibinfo {author} {\bibfnamefont
  {D.}~\bibnamefont {Stump}}, \bibinfo {author} {\bibfnamefont {W.-K.}\
  \bibnamefont {Tung}}, \ and\ \bibinfo {author} {\bibfnamefont {C.~P.}\
  \bibnamefont {Yuan}},\ }\href {\doibase 10.1103/PhysRevD.78.013004}
  {\bibfield  {journal} {\bibinfo  {journal} {Phys. Rev.}\ }\textbf {\bibinfo
  {volume} {D78}},\ \bibinfo {pages} {013004} (\bibinfo {year} {2008})},\
  \Eprint {http://arxiv.org/abs/0802.0007} {arXiv:0802.0007 [hep-ph]}
  \BibitemShut {NoStop}%
\bibitem [{\citenamefont {Chollet}\ \emph {et~al.}()\citenamefont {Chollet}
  \emph {et~al.}}]{chollet2015keras}%
  \BibitemOpen
  \bibfield  {author} {\bibinfo {author} {\bibfnamefont {F.}~\bibnamefont
  {Chollet}} \emph {et~al.},\ }\href@noop {} {\enquote {\bibinfo {title}
  {Keras},}\ }\bibinfo {howpublished}
  {\url{https://github.com/fchollet/keras}}\BibitemShut {NoStop}%
\bibitem [{\citenamefont {Abadi}\ \emph {et~al.}(2016)\citenamefont {Abadi}
  \emph {et~al.}}]{Abadi:2016:TSL:3026877.3026899}%
  \BibitemOpen
  \bibfield  {author} {\bibinfo {author} {\bibfnamefont {M.}~\bibnamefont
  {Abadi}} \emph {et~al.},\ }in\ \href
  {http://dl.acm.org/citation.cfm?id=3026877.3026899} {\emph {\bibinfo
  {booktitle} {Proceedings of the 12th USENIX Conference on Operating Systems
  Design and Implementation}}},\ \bibinfo {series and number} {OSDI'16}\
  (\bibinfo  {publisher} {USENIX Association},\ \bibinfo {address} {Berkeley,
  CA, USA},\ \bibinfo {year} {2016})\ pp.\ \bibinfo {pages}
  {265--283}\BibitemShut {NoStop}%
\bibitem [{\citenamefont {Nickolls}\ \emph {et~al.}(2008)\citenamefont
  {Nickolls}, \citenamefont {Buck}, \citenamefont {Garland},\ and\
  \citenamefont {Skadron}}]{Nickolls:2008:SPP:1365490.1365500}%
  \BibitemOpen
  \bibfield  {author} {\bibinfo {author} {\bibfnamefont {J.}~\bibnamefont
  {Nickolls}}, \bibinfo {author} {\bibfnamefont {I.}~\bibnamefont {Buck}},
  \bibinfo {author} {\bibfnamefont {M.}~\bibnamefont {Garland}}, \ and\
  \bibinfo {author} {\bibfnamefont {K.}~\bibnamefont {Skadron}},\ }\href
  {\doibase 10.1145/1365490.1365500} {\bibfield  {journal} {\bibinfo  {journal}
  {Queue}\ }\textbf {\bibinfo {volume} {6}},\ \bibinfo {pages} {40} (\bibinfo
  {year} {2008})}\BibitemShut {NoStop}%
\bibitem [{\citenamefont {{Chetlur}}\ \emph {et~al.}()\citenamefont
  {{Chetlur}}, \citenamefont {{Woolley}}, \citenamefont {{Vandermersch}},
  \citenamefont {{Cohen}}, \citenamefont {{Tran}}, \citenamefont
  {{Catanzaro}},\ and\ \citenamefont {{Shelhamer}}}]{cuDNN}%
  \BibitemOpen
  \bibfield  {author} {\bibinfo {author} {\bibfnamefont {S.}~\bibnamefont
  {{Chetlur}}}, \bibinfo {author} {\bibfnamefont {C.}~\bibnamefont
  {{Woolley}}}, \bibinfo {author} {\bibfnamefont {P.}~\bibnamefont
  {{Vandermersch}}}, \bibinfo {author} {\bibfnamefont {J.}~\bibnamefont
  {{Cohen}}}, \bibinfo {author} {\bibfnamefont {J.}~\bibnamefont {{Tran}}},
  \bibinfo {author} {\bibfnamefont {B.}~\bibnamefont {{Catanzaro}}}, \ and\
  \bibinfo {author} {\bibfnamefont {E.}~\bibnamefont {{Shelhamer}}},\
  }\href@noop {} {\bibinfo  {journal} {arXiv:1410.0759}\ }\BibitemShut
  {NoStop}%
\bibitem [{\citenamefont {{Kingma}}\ and\ \citenamefont {{Ba}}(2014)}]{adam}%
  \BibitemOpen
\bibfield  {journal} {  }\bibfield  {author} {\bibinfo {author} {\bibfnamefont
  {D.~P.}\ \bibnamefont {{Kingma}}}\ and\ \bibinfo {author} {\bibfnamefont
  {J.}~\bibnamefont {{Ba}}},\ }\href@noop {} {\bibfield  {journal} {\bibinfo
  {journal} {arXiv:1412.6980}\ } (\bibinfo {year} {2014})}\BibitemShut
  {NoStop}%
\bibitem [{\citenamefont {{Bengio}}(2012)}]{earlystopping}%
  \BibitemOpen
  \bibfield  {author} {\bibinfo {author} {\bibfnamefont {Y.}~\bibnamefont
  {{Bengio}}},\ }\href@noop {} {\bibfield  {journal} {\bibinfo  {journal}
  {arXiv:1206.5533}\ } (\bibinfo {year} {2012})}\BibitemShut {NoStop}%
\bibitem [{\citenamefont {{Klambauer}}\ \emph {et~al.}(2017)\citenamefont
  {{Klambauer}}, \citenamefont {{Unterthiner}}, \citenamefont {{Mayr}},\ and\
  \citenamefont {{Hochreiter}}}]{selu}%
  \BibitemOpen
  \bibfield  {author} {\bibinfo {author} {\bibfnamefont {G.}~\bibnamefont
  {{Klambauer}}}, \bibinfo {author} {\bibfnamefont {T.}~\bibnamefont
  {{Unterthiner}}}, \bibinfo {author} {\bibfnamefont {A.}~\bibnamefont
  {{Mayr}}}, \ and\ \bibinfo {author} {\bibfnamefont {S.}~\bibnamefont
  {{Hochreiter}}},\ }\href@noop {} {\bibfield  {journal} {\bibinfo  {journal}
  {arXiv:1706.02515}\ } (\bibinfo {year} {2017})}\BibitemShut {NoStop}%
\bibitem [{\citenamefont {Buckley}(2015)}]{Buckley:2013jua}%
  \BibitemOpen
  \bibfield  {author} {\bibinfo {author} {\bibfnamefont {A.}~\bibnamefont
  {Buckley}},\ }\href {\doibase 10.1140/epjc/s10052-015-3638-8} {\bibfield
  {journal} {\bibinfo  {journal} {Eur. Phys. J.}\ }\textbf {\bibinfo {volume}
  {C75}},\ \bibinfo {pages} {467} (\bibinfo {year} {2015})},\ \Eprint
  {http://arxiv.org/abs/1305.4194} {arXiv:1305.4194 [hep-ph]} \BibitemShut
  {NoStop}%
\bibitem [{\citenamefont {Allanach}\ \emph {et~al.}(2009)\citenamefont
  {Allanach} \emph {et~al.}}]{Allanach:2008qq}%
  \BibitemOpen
  \bibfield  {author} {\bibinfo {author} {\bibfnamefont {B.~C.}\ \bibnamefont
  {Allanach}} \emph {et~al.},\ }\href {\doibase 10.1016/j.cpc.2008.08.004}
  {\bibfield  {journal} {\bibinfo  {journal} {Comput. Phys. Commun.}\ }\textbf
  {\bibinfo {volume} {180}},\ \bibinfo {pages} {8} (\bibinfo {year} {2009})},\
  \Eprint {http://arxiv.org/abs/0801.0045} {arXiv:0801.0045 [hep-ph]}
  \BibitemShut {NoStop}%
\bibitem [{\citenamefont {Otten}\ \emph {et~al.}()\citenamefont {Otten} \emph
  {et~al.}}]{DeepXS}%
  \BibitemOpen
  \bibfield  {author} {\bibinfo {author} {\bibfnamefont {S.}~\bibnamefont
  {Otten}} \emph {et~al.},\ }\href@noop {} {\enquote {\bibinfo {title}
  {{DeepXS}},}\ }\bibinfo {howpublished}
  {\url{https://github.com/SydneyOtten/DeepXS}}\BibitemShut {NoStop}%
\bibitem [{\citenamefont {Caron}\ \emph {et~al.}(2017)\citenamefont {Caron},
  \citenamefont {Kim}, \citenamefont {Rolbiecki}, \citenamefont {Ruiz~de
  Austri},\ and\ \citenamefont {Stienen}}]{Caron:2016hib}%
  \BibitemOpen
  \bibfield  {author} {\bibinfo {author} {\bibfnamefont {S.}~\bibnamefont
  {Caron}}, \bibinfo {author} {\bibfnamefont {J.~S.}\ \bibnamefont {Kim}},
  \bibinfo {author} {\bibfnamefont {K.}~\bibnamefont {Rolbiecki}}, \bibinfo
  {author} {\bibfnamefont {R.}~\bibnamefont {Ruiz~de Austri}}, \ and\ \bibinfo
  {author} {\bibfnamefont {B.}~\bibnamefont {Stienen}},\ }\href {\doibase
  10.1140/epjc/s10052-017-4814-9} {\bibfield  {journal} {\bibinfo  {journal}
  {Eur. Phys. J.}\ }\textbf {\bibinfo {volume} {C77}},\ \bibinfo {pages} {257}
  (\bibinfo {year} {2017})},\ \Eprint {http://arxiv.org/abs/1605.02797}
  {arXiv:1605.02797 [hep-ph]} \BibitemShut {NoStop}%
\end{thebibliography}%

\end{document}